\documentclass[preprint,12pt]{aastex}
\usepackage{emulateapj5}

\newcommand{\etal}{{\it et al.}}
\newcommand{\chisq}{\mbox{$\chi^{2}$}}
\newcommand{\kms}{km\,s$^{-1}$}

\newcommand{\kmsyr}{km\,s$^{-1}$ yr$^{-1}$}
\newcommand{\muasyr}{$\mu$as\,yr$^{-1}$}

\newcommand{\percmcm}{cm$^{-2}$}
\newcommand{\percmcmcm}{cm$^{-3}$}
\newcommand{\msun}{\mbox{M$_{\odot}$}}
\newcommand{\dmass}{\mbox{$M_{D}$}}
\newcommand{\mdot}{\mbox{$\dot{M}$}}

\newcommand{\smy}{\mbox{M$_{\odot}$\,yr$^{-1}$}}

\newcommand{\ha}{\mbox{H$\alpha$}}
\newcommand{\muas}{\mbox{$\mu$as}}
\newcommand{\bq}{\begin{equation}}
\newcommand{\eq}{\end{equation}}

\newcommand{\ergs}{\mbox{erg\,s$^{-1}$}}
\newcommand{\rg}{\mbox{r$_{g}$}}
\newcommand{\rtrans}{\mbox{$r_{tr}$}}

\newcommand{\degree}{\mbox{$\circ$}}
\newcommand{\rhoc}{\mbox{$\rho_{c}$}}

\newcommand{\didr}{\mbox{$di/dr$}}
\newcommand{\didrp}{\mbox{$di/dr'$}}
\newcommand{\dmassi}{\mbox{$M_{D_{i}}$}}
\newcommand{\smpc}{\mbox{M$_{\odot}$\,pc$^{-3}$}}
\newcommand{\pang}{\mbox{$\phi$}}
\newcommand{\xo}{\mbox{$x_{o}$}}
\newcommand{\xop}{\mbox{$x'_{o}$}}
\newcommand{\xof}{\mbox{$x^{f}_{o}$}}
\newcommand{\xxi}{\mbox{$x_{i}$}}
\newcommand{\yo}{\mbox{$y_{o}$}}
\newcommand{\yi}{\mbox{$y_{i}$}}
\newcommand{\vo}{\mbox{$v_{o}$}}
\newcommand{\vof}{\mbox{$v^{f}_{o}$}}
\newcommand{\mo}{\mbox{$M_{o}$}}
\newcommand{\mass}{\mbox{${\cal M}$}}
\newcommand{\rc}{\mbox{$r_{c}$}}
\newcommand{\ri}{\mbox{$r_{i}$}}
\newcommand{\ro}{\mbox{$r_{o}$}}
\newcommand{\rs}{\mbox{$r_{s}$}}
\newcommand{\is}{\mbox{$i_{s}$}}	
\newcommand{\io}{\mbox{$i_{o}$}}
\newcommand{\ii}{\mbox{$i_{i}$}}
\newcommand{\vmeasi}{\mbox{$v_{i}$}}
\newcommand{\vmodi}{\mbox{$v^{\prime}_{los_{i}}$}}
\newcommand{\vreli}{\mbox{$v_{rel_{i}}$}}
\newcommand{\vlosi}{\mbox{$v_{los_{i}}$}}
\newcommand{\dtheta}{\mbox{$\delta\vec{\theta}_{r}$}}
\newcommand{\domega}{\mbox{$\Delta\omega$}}
\newcommand{\vlos}{\mbox{$v_{los}$}}
\newcommand{\vroti}{\mbox{$v_{rot_{i}}$}}
\newcommand{\vrot}{\mbox{$v_{rot}$}}
\newcommand{\sigyp}{\mbox{$\sigma^{\prime}_{y}$}}
\newcommand{\sigmath}{\mbox{$\sigma_{\theta}$}}
\newcommand{\sigr}{\mbox{$\sigma_{r}$}}
\newcommand{\sigx}{\mbox{$\sigma_{x}$}}

\newcommand{\sigv}{\mbox{$\sigma_{v}$}}
\newcommand{\sigxi}{\mbox{$\sigma_{x_{i}}$}}
\newcommand{\sigyi}{\mbox{$\sigma_{y_{i}}$}}
\newcommand{\rmin}{\mbox{$r_{min}$}}

\newcommand{\sound}{\mbox{$c_{s}$}}
\newcommand{\taumax}{\mbox{$\tau_{max}$}}
\newcommand{\microns}{\mbox{$\mu$m}}

\newcommand{\sigmavi}{\mbox{$\sigma_{v_{i}}$}}
\newcommand{\ns}{\mbox{$N_{24}$}}
\newcommand{\mdotmf}{\mbox{$\dot{M}_{-5}$}}
\newcommand{\mdota}{\mbox{$\dot{M}\alpha^{-1}$}}
\newcommand{\masse}{\mbox{$M_{8}$}}
\newcommand{\rpc}{\mbox{$r_{pc}$}}
\newcommand{\rcr}{\mbox{$r_{cr}$}}
\newcommand{\tvar}{\mbox{$t_{var}$}}
\newcommand{\rx}{\mbox{$r_{x}$}}
\newcommand{\lum}{\mbox{$L_{41}$}}
\newcommand{\lbol}{\mbox{$L_{bol}$}}
\newcommand{\ledd}{\mbox{$L_{E}$}}
\newcommand{\css}{\mbox{$c_{s_{7}}$}}
\newcommand{\cs}{\mbox{$c_{s}$}}
\newcommand{\rpm}{\mbox{mas$^{-1}$}}
\newcommand{\router}{\mbox{$r_{out}$}}
\newcommand{\tm}{\mbox{$T_{m}$}}
\newcommand{\rhom}{\mbox{$\rho_{m}$}}
\newcommand{\rhomid}{\mbox{$\rho_{mid}$}}

\newcommand{\tv}{\mbox{$t_{v}$}}
\newcommand{\thetad}{\mbox{$\theta$}}

\newcommand{\nhat}{\mbox{$\hat{n}$}}

\newcommand{\rhat}{\mbox{$\hat{r}$}}

\begin{document}
 
\title{The Geometry of and Mass Accretion Rate through the Maser Accretion Disk in 
NGC 4258}
 
\author{
J. R. Herrnstein\altaffilmark{1},
J. M. Moran\altaffilmark{2},
L. J. Greenhill\altaffilmark{2}, and
Adam S. Trotter\altaffilmark{3}}
\altaffiltext{1}{Renaissance Technologies, Corp., 600 Rt. 25A, E. Setauket, NY 11733, jrh@rentec.com}
\altaffiltext{2}{Harvard-Smithsonian Center for Astrophysics, Mail Stop 42, 
60 Garden Street, Cambridge, MA 02138}
\altaffiltext{3}{Department of Physics and Astronomy, University of
North Carolina at Chapel Hill, Chapel Hill, NC 27599-3255}

\slugcomment{accepted for publication in ApJ}

\begin{abstract}
A maximum likelihood analysis of the NGC~4258 maser positions and
velocities reveals a $\sim2\sigma$ deviation from Keplerian motion in
the projected rotation curve of the high-velocity features,
corresponding to a $\sim9$\,\kms, or 0.8\%, flattening of the LOS
velocities with respect to Keplerian motion over the range of the
high-velocity masers.  While there are a number of potential
explanations for this flattening, we argue for pure Keplerian rotation
in an inclination-warped disk based on the ability of this model to
explain a number of otherwise puzzling features of the system. A radial
gradient in the disk inclination of 0.034\,\rpm\ is not only
consistent with the observed rotation curve, but it generates a bowl
along the near edge of the disk that naturally explains the otherwise
puzzling narrow spread in the declinations of the systemic masers. It
also explains the existence and location of an apparently recurring
flare amongst the systemic masers.  There is no significant evidence
for non-Keplerian rotation in the inclination-warped disk.  An
additional implication of the inclination warp is that the disk rises
in front of and obscures the central engine at a disk radius of about
8.3\,mas, or 0.29\,pc. By comparing the observed X-ray column to
conditions in the disk at this radius, we argue the disk must be
atomic at 0.29\,pc. Hence we conclude that the molecular-to-atomic
transition occurs just beyond the outermost maser at 0.28\,pc and from
this we infer an accretion rate of $\sim 10^{-4}\alpha$\,\smy, where
$\alpha$ ($\la1$) is the standard dimensionless parameterization of
the kinematic viscosity.  Our model suggests that most of the observed
X-ray column arises in the warped accretion disk at 0.29\,pc, and that
the maser emission is truncated at large radii predominantly as a
result of the molecular-to-atomic phase transition originally proposed
by Neufeld \& Maloney (1995).  The inferred accretion rate is
consistent with the jet-dominated accretion models of Yuan \etal\
(2002).
\end{abstract}



\section{Introduction}

NGC 4258 (M106) is a prominent barred spiral galaxy (SABbc) located at
a distance of 7.2\,Mpc (Herrnstein \etal\ 1998b). It is one of 12
galaxies included in Seyfert's initial catalog of active galaxies
(Seyfert 1943) and it has been studied extensively over a broad range
of frequencies and resolutions.  On large scales, NGC 4258 is
dominated by a pair of anomalous spiral arms positioned between the
stellar arms and symmetric with respect to the nucleus. The nonthermal
emission from these arms spans many decades in frequency and is
probably energized in shocks formed at the interface between the ISM
and matter ejected from the nucleus (van der Kruit, Oort, \& Mathewson
1972; Ford \etal\ 1986). Higher resolution \ha\ and VLA radio
continuum images reveal that the anomalous arms are fed by
north-south, multi-strand, twisted jets arising in the active nucleus
of NGC 4258 (Cecil, Wilson, \& Tully 1992; Cecil, Wilson, \& De Pree
1995). The jets bifurcate and bend sharply about 2.6 kpc from the
central engine, presumably as a result of interactions with especially
dense regions in the ISM (Martin \etal\ 1989).

Further evidence for nuclear activity comes from IR, optical, and
X-ray observations. Chary \& Becklin (1997) detect a 4.5 mJy compact
central source at 2.2\,\microns\ whose color is consistent with that
of an extincted nonthermal source. The near IR emission is unresolved
by the $0''.6$ (21\,pc) beam.  Chary \etal\ (2000) present higher
resolution images between 1 and 18\,\microns, and report nuclear
emission unresolved by the $0''.2$ beam with a spectral index, $s$, of
$1.4\pm0.1$ ($f_{\nu}=\nu^{-s}$) and a 1--20\,\microns\ luminosity of
$\sim8\times10^{41}$\,\ergs.  At optical wavelengths, Wilkes \etal\
(1995) detect faint blue continuum emission and broadened
($\sim2500$\,\kms, FWHM) emission in polarized light. This is
presumably the scattered emission from an obscured Seyfert 1 nucleus,
and NGC~4258 is characterized as a hidden broad line region (HBLR)
Seyfert 2 AGN.  Wilkes \etal\ infer a spectral index of $1.1\pm0.2$
and an optical luminosity of $5.5\times10^{42}$\,\ergs\ or
$5.5\times10^{40}$\,\ergs, depending on whether the scattering screen
is composed of electrons or dust.  Finally, NGC~4258 has been observed
by several X-ray telescopes over the last decade (see Table~2 of
Fruscione \etal\ (2005) for a summary).  The most recent XMM-{\it
Newton} data of Fruscione \etal\ (2005) suggest a power law spectrum
with a neutral hydrogen absorbing column of
$8.6-13.2\times10^{22}$\,\percmcm\ and an absorption-corrected 2--10
keV luminosity of $5.4-9.1\times10^{40}$\,\ergs.  Taken together,
these data imply a bolometric luminosity (\lbol) of
$\sim10^{42}$\,\ergs.

While informative, none of these data actually resolve the NGC 4258
central engine on sub-parsec-scales, where the details of the
accretion processes powering the AGN are likely to be uncovered.
Fortunately, NGC 4258 contains a remarkable nuclear water maser, and
VLBI observations of these dense molecular gas tracers provide a
unique window into the sub-parsec-scale structure of the AGN.  The NGC
4258 water maser was initially discovered by Claussen, Heiligman, \&
Lo (1984) and was shown to be confined to the inner 3 pc of the galaxy
shortly thereafter (Claussen \& Lo 1986).  The unexpected discovery of
maser emission offset $\pm1000$\,\kms\ on either side of the brighter
emission at the galaxy systemic velocity of 470\,\kms\ (Nakai, Inoue,
\& Miyoshi 1993) suggested the possibility of a rapidly rotating
structure. Greenhill \etal\ (1995a) showed that the systemic masers
lie along a narrow, east-west structure spanning 4 mas and reported a
striking linear relationship between position and line-of-sight (LOS)
velocity that, in the context of disk models, suggested either solid
body rotation or confinement to a narrow range of radii.  In parallel
with these developments, two teams were completing comprehensive,
10-year, single-dish monitoring programs of the systemic maser
emission. Both Haschick, Baan, \& Peng (1994) and Greenhill \etal\
(1995b) reported a secular drift in the systemic maser LOS velocities
of about 9\,\kmsyr. In addition, Greenhill \etal\ made the preliminary
conclusion that the high-velocity masers show no acceleration.  Watson
\& Wallin (1994) were the first to propose a self-consistent disk
model in which the systemic masers lie along a narrow annulus at the
near edge of an edge-on disk and the high-velocity masers are near the
diameter defined by the intersection of the disk with the plane of the
sky (the midline).

\begin{figure}[p]          
\plotone{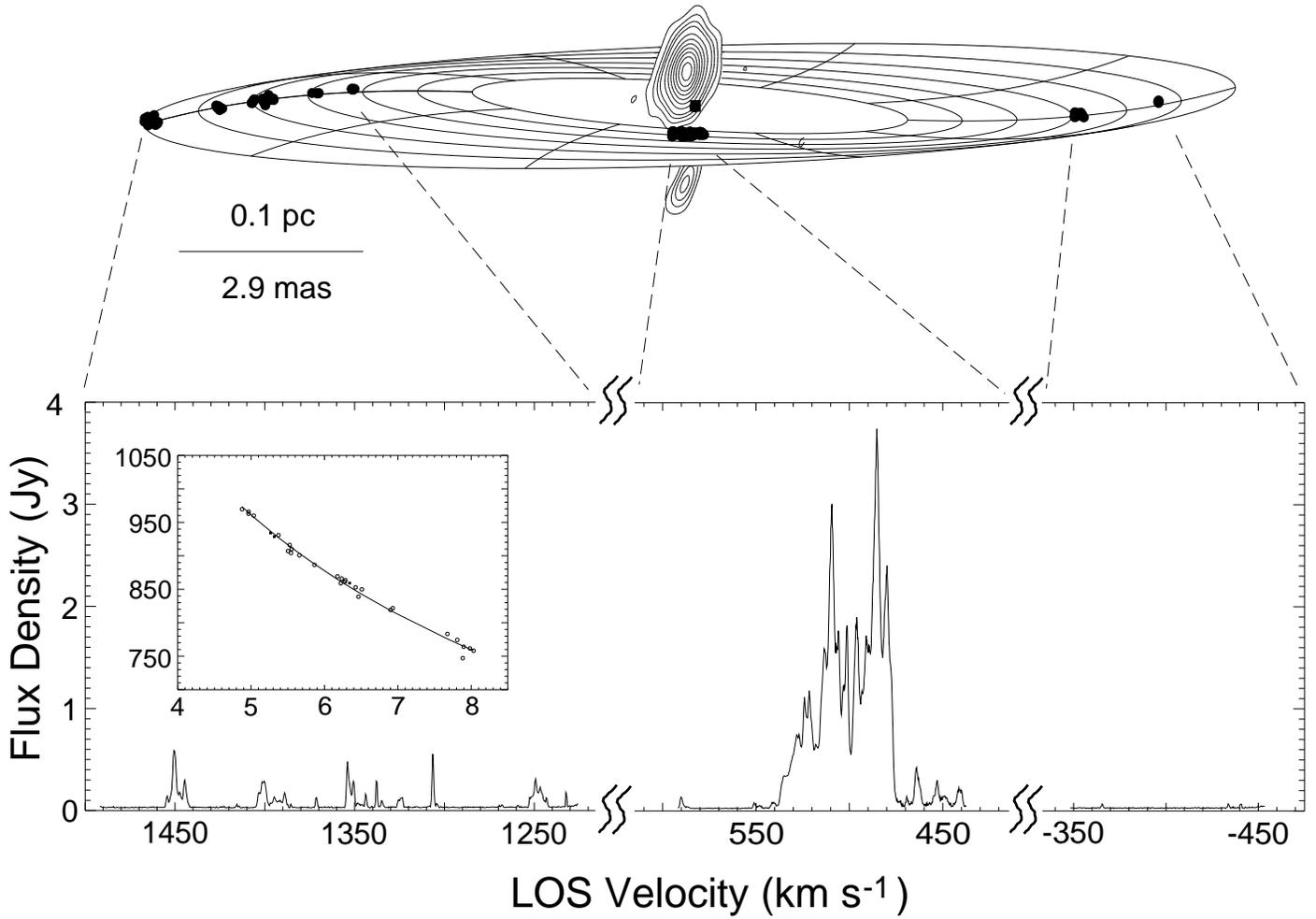}
\caption {{\it Top:} Warped-disk model with masers and continuum
superposed.  {\it Bottom:} Total power spectrum of the NGC 4258 maser,
with best-fitting Keplerian rotation curve.  Maser data are from VLBA
epoch BM36b, continuum data are from epochs BH25a-c (see Table~1).}
\label{fg:orig}
\end{figure}

The VLBI image produced by Miyoshi \etal\ (1995) provides the first
unequivocal demonstration that the masers trace a well-organized
rotating disk (see Figure~\ref{fg:orig}). The bulk of high-velocity
masers extends from 0.17 to 0.28 pc on either side of the systemic
masers and apparently obeys a Keplerian rotation law. (Note that an
isolated feature has been detected recently at 0.12~pc (Modjaz et
al. 2005).)  The rotation curve requires a central binding mass of
$3.8\times10^{7}$\,\msun\ and a central density of at least
$4\times10^{9}$\,\smpc.  This central mass, together with the systemic
maser position-velocity gradient and LOS accelerations, indicates the
systemic masers occupy a narrow annulus 0.14 pc from the central
engine and spanning $8^{\circ}$ in disk azimuth along the near edge of
the disk. At 1100\,\kms\ it takes individual masers about 10 years to
traverse the entire envelope of systemic emission.  The systemic
masers lie about 0.02 pc south of the disk center as determined from
the high-velocity masers, requiring that the disk be tipped down about
$8^{\circ}$ from edge on. The lack of structure in the vertical
positions of the systemic masers implies that the aspect ratio of the
disk is less than about 0.2\% and that the portion of the disk traced
by the masers is extremely thin (Moran \etal\ 1995). Because the
high-velocity masers are confined to within several degrees of the
midline by their small observed accelerations (Bragg \etal\ 2000), the
systematic curvature in the declinations of these features implies
that the disk is warped. The data are well fit by a gradual
$8^{\degree}$ change in the position angle of the disk between 0.17
and 0.28 pc (Herrnstein 1997).  Herrnstein, Greenhill, \& Moran (1996)
use this warp to explain the relative faintness of the blue-shifted
emission.  An ongoing multi-epoch VLBA monitoring program has detected
$\sim30$\,\muasyr\ proper motions of the systemic masers as they sweep
in front of the background jet at 1100\,\kms.  In addition to
confirming the disk geometry, these proper motions, together with the
accelerations, provide a geometric distance to the galaxy (Herrnstein
\etal\ 1998b).

High-frequency VLBA continuum observations have uncovered a compact
jet elongated along the rotation axis of the disk, and well-aligned
with the large scale jets at the base of the anomalous arms
(Herrnstein \etal\ 1997).  The $\sim0.01$ pc offset between the jet
emission and the central engine is consistent with optical depth
effects along a mildly relativistic jet (Blandford \& Konigl 1979).
The southern jet is 5--10 times weaker than the northern jet, and this
is probably due to thermal absorption in the disk.  Furthermore, while
the high-velocity masers presumably amplify their own spontaneous
emission (the velocity gain paths being longest along the midline),
the systemic masers most likely amplify the southern jet, which lies
directly behind them.  VLBA observations provide a 3-$\sigma$ upper
limit of 220\,$\mu$Jy on any 22 GHz continuum emission coincident with
the central engine (Herrnstein \etal\ 1998a).

Bragg \etal\ (2000) report the  detection with the VLA of a flare in the
systemic spectrum in 1996, at a velocity of 492.6\,\kms\ LSR.  The flux
density of this flare was 20 Jy, a factor of 4--5 brighter
than the typical peak in the systemic emission.  The flare maximum was
observed on 29 March 1996.  There is some evidence for above-normal
emission at the same velocity in the preceding observation of February
26.  However, the observation following the flare peak, on 10 May
1996, shows no evidence of the flare.  From these data, an upper limit
on the flare lifetime of $\sim75$ days is inferred.

While the warped-disk model derived from the positions, LOS
velocities, motions, and accelerations of the NGC 4258 masers is
compelling in its completeness and self-consistency, several important
questions remain unanswered. In particular, in spite of all that is
known about this AGN central engine, the accretion rate, \mdot,
through the disk is still a matter of speculation.  The central mass
together with the estimated bolometric luminosity indicate a highly
sub-Eddington system, $\lbol\sim10^{-3}-10^{-4}$\,\ledd, the Eddington
luminosity corresponding to the central mass.  In fact, NGC~4258 is
quite characteristic of the ubiquitous class of relatively faint AGN
known collectively as low luminosity AGN (LLAGN).  A determination of
\mdot\ might provide a means to discriminate amongst the various
accretion modes that have been proposed in NGC~4258, and shed some
light on accretion processes in LLAGN in general. Unfortunately, the
infall motions of the gas cannot be measured directly in NGC~4258, and
estimates for \mdot\ have been indirect and insufficiently precise to
constrain the accretion models.

A second category of unanswered questions concerns the geometry of the
disk itself. The warping in the position angle of the disk is
immediately obvious because the disk mid-line is well-sampled by the
high-velocity masers.  A corresponding warping in the inclination
($i$) of the disk (i.e. $i=i(r)$) is very likely, as the alternative
would imply that we observe the disk from a very special
perspective. Unfortunately, the sparse sampling of the disk means that
observations are not terribly sensitive to an inclination warp
(Herrnstein, Greenhill, \& Moran 1996).  There are also questions
concerning the precise locations of the masers in the disk.  It has
long been appreciated that the high-velocity masers probably occupy
the midline because this is where the velocity gain paths are longest.
Furthermore, Neufeld \& Maloney (1995) explain the outer edge of the
maser emission in the context of a molecular-to-atomic phase
transition. However, there is no compelling explanation for why the
systemic masers are so narrowly clustered in declination and
apparently in radius. Nor is it understood why the high-velocity
features possess an apparently abrupt inner edge at 0.17\,pc on both
the red-shifted and blue-shifted sides of the disk.

In this paper we use the shape of the high-velocity maser rotation
curve to explore the three-dimensional geometry of the disk between
0.14 and 0.28\,pc. We find that an inclination warp consistent with
the rotation curve also naturally explains the morphology of the
systemic masers.  In addition, we report the discovery of a second
flare at the same velocity as the first, and show how the inclination
warp can explain this apparently recurring feature of the system.
An additional implication of the inclination warp is that the disk itself
obscures the central engine X-ray emission. By comparing the observed
X-ray column to the predicted disk column, we are able to place tight
constraints on the accretion rate through the maser disk. 

Section~2 discusses the data and observations utilized in the
analysis. Section~3 presents the rotation curve fits.  Section~4
demonstrates that the inclination-warp model provides a compelling
explanation for the otherwise puzzling geometry of the systemic
masers.  In Section~5 we consider various implications of the
inclination warp, focusing on the X-ray column and the accretion rate.
Section~6 presents concluding remarks.

\section{Observations and Data}

The purpose of this analysis is to explore deviations from pure
Keplerian rotation in the projected rotation curve of the
high-velocity masers.  Isolating such deviations requires careful
statistical analysis, since the precise shape of the rotation curve
correlates strongly with several additional fitted parameters, most
notably the position of the disk center.  We attempt to minimize
statistical uncertainties in our results by combining a number of
datasets, listed in Table~1.  The first four epochs (designated 'L' in
the table) were designed to deliver synthesis maps over the entire
range of maser emission.  The principle objective of the 'C' epochs,
on the other hand, was high-sensitivity, \muas\ relative-position
images of the central continuum, and they yielded synthesis maps for
the systemic masers only. Each of the epochs included the entire
VLBA\footnote{The National Radio Astronomy Observatory is a facility
of the National Science Foundation operated under cooperative
agreement by Associated Universities, Inc.} plus the phased VLA
acting as a single large aperture.  The NRAO 140-foot and Effelsberg
100-meter were included in several of the experiments. The relatively
narrow bandwidth provided by the VLA complicated the observations
considerably.  While the masers span about 150\,MHz, the VLA is
limited to a 50\,MHz bandwidth in each polarization. Thus, in each
observation one polarization was centered on the systemic masers while
the other polarization alternated between either the red-shifted and
blue-shifted masers ('L'), or bands devoid of maser emission, well
outside of the highest-velocity masers ('C'). In the 'L' configuration
we observed a total of four 8-MHz bands, and correlated 512 spectral
channels across each band giving a spectral resolution of about
0.22\,\kms.  The 'C' configuration utilized eight 8-MHz bands and 256
channels, giving a spectral resolution of 0.44\,\kms\ across the
systemic maser emission. In both frequency setups, the interferometer
was stabilized by applying the phase of a single strong systemic maser
to the entire spectrum. While this results in the loss of all absolute
position information, it provides relative positions of a small
fraction of the 500\,\muas\ beam.

Our need to phase reference across the full 150\,MHz spectrum, and
across different polarizations, introduced a number of complications
into the data reduction (Herrnstein 1997). Errors, \dtheta, in the
absolute position of the reference maser lead to second order errors
in maser positions of $[\domega/\omega]\dtheta$, where \domega\ is the
offset in frequency between the reference and imaged maser, and
$\omega$ is the observing frequency.  This amounts to
3.5\,\muas\,mas$^{-1}$ at 22 GHz, across the 75 MHz spread of the NGC
4258 maser.  In this analysis, we reference all the data to the
systemic maser at 510\,\kms\ using the position derived by Herrnstein
(1997) via a fringe rate analysis on four VLBA epochs:

\begin{center}
\noindent $\alpha = 12^h~18^m~57.5046^s\pm0.0003$  (J2000)\\
$\delta = 47^\circ~18'~14.303''\pm0.003$  (J2000)
\end{center}

\noindent The $\sim5$ mas uncertainty in the reference maser
corresponds to a negligible 18~\muas\ uncertainty across the full
extent of NGC 4258.  In general, broad bandwidth phase referencing
leads to a heightened sensitivity to all forms of delay-like errors.
As a result, delay calibrators were observed at 10--20 minute
intervals throughout the experiments. These calibrators were also used
to calibrate the relative phase offsets between the two polarizations,
a critical step in the calibration process given our need to phase
reference across polarizations.  The VLA was phased-up on a nearby,
strong calibrator prior to each NGC 4258 scan, and typically
maintained phase coherence at the 80--90\% level across the 10--15
minute NGC 4258 scans. The relative amount of time on phase-up sources
and on NGC 4258 at the VLA was adjusted in real time according to
changes in weather conditions and phase stability.

Each spectral channel was imaged using the standard image
deconvolution techniques provided in AIPS. Thermal noises of between
10 and 20 mJy per beam per spectral channel were achieved. All
features stronger than $2\sigma$ that occurred in two adjacent
frequency planes within a beam of one another were considered
significant.  The centroided positions of these features were
determined using a Gaussian fitting routine in AIPS.  Figure
\ref{fg:orig} includes a spectrum derived from the image cubes in this
way. The positions and velocities of individual masers were derived
from this spectrum using a crude feature decomposition algorithm.
Each peak in the interferometric spectrum was assumed to correspond to
a distinct maser feature.  The peaks were determined from numerical
first and second derivatives of the spectrum, and the positions and
velocities of the resulting maser features are the flux-weighted
averages around these peaks.  The result is 25--35 systemic masers,
10--20 red-shifted masers, and 2--4 blue-shifted masers, per
epoch. Each epoch was referenced to the same point in the systemic
maser emission envelope to ensure consistency.  Continuum maps were
generated by averaging all maser-free channels.

\subsection{Single-Dish Monitoring of the Systemic Spectrum}

In hopes of detecting additional systemic flares, the low-velocity
maser spectrum of NGC4258 was monitored from 1997 January 27 to 1997
June 24 with the 37-meter diameter antenna of the Haystack
Observatory.  Eleven epochs were observed.  Position-switched total
power spectra were calibrated using a system temperature referenced to
outside the atmosphere and corrected for elevation dependent antenna
gain using a standard model for the antenna.  For each observing
track, we observed continuum sources 4C39.25 and 1308+326 to calibrate
offsets in antenna pointing. To correct for pointing errors within one
hour of transit (elevations $\ga 78^\circ$), we sampled the peak line
intensity every $\sim 10$ minutes and rescaled the spectra to achieve
a level response.  The calibration of the spectrum compiled for each
epoch is accurate to $\sim20\%$.

We used a 17.8 MHz bandwidth in one circular polarization and obtained
8192 channel spectra to which we applied a boxcar convolving function
to obtain a channel spacing of 21.7 kHz (0.293\,\kms).  The rms noise
after removal of a spectral baseline was 0.1 to 0.2 Jy per channel for
the 11 epochs.

\section{The High-Velocity Rotation Curve}

\subsection{The Basic Model}

The shape of the high-velocity rotation curve is explored using a
\chisq\ minimization analysis comparing model LOS velocities to
measured velocities.  The details are described in this section. In
the interest of generality, we revert to angular coordinates for the
maser positions. Thus, the systemic masers are clustered about a disk
radius of 3.9 mas, the high-velocity masers extend from 4.5 to 8.1
mas, and at 7.2\,Mpc, one mas equals 0.035 pc.  The basic thin-disk
model assumes that the masers obey simple Keplerian motion: 
\bq
\vroti=\left[\frac{G\mass}{\sqrt{(\xxi-\xo)^{2}+(\yi-\yo)^{2}}}\right]^{1/2}.
\label{eq:kepler}
\eq 
Here, \xxi\ and \yi\ are the angular sky coordinates of the
$i^{th}$ maser, $G$ is the gravitational constant, \mass\ is the
central mass (\mo) over the distance ($d$), and \xo\ and \yo\ specify
the disk center. For masers assumed to be along the midline, this
rotation velocity is related to the LOS velocity (\vlosi)
according to 
\bq 
\vlosi=\vroti\sin(i)+\vo,
\label{eq:dumb1}
\eq
where $i$ and \vo\ are the inclination ($90^{\circ}$ corresponding to
edge-on) and systemic velocity of the disk, respectively.

Rotation curve models are assessed by comparing \vlosi\ to the
observed maser LOS velocities. Special and general relativistic
effects play a significant role in the transformation from observed
frequency to LOS velocity for the masers. Since the relativistic
corrections are model-dependent, we choose to remove them from the
model LOS velocities instead of including them in the observed
velocities.  Hence, the relativity-corrected model LOS velocity
(\vmodi) is given by (see, for example, Herrnstein 1997 for details)
\bq 
\vmodi=\vlosi+\vreli
\label{eq:dumb11}
\eq
where
\bq
\vreli\simeq\frac{1}{2c}\left[\vlosi^{2}+\vroti^{2}\cos^{2}(i)+2\vroti^{2}\right].
\label{eq:model1}
\eq
These model velocities are compared to the observed velocities (\vmeasi)
as derived by the classical expression for the Doppler shift (optical
definition):
\bq
\vmeasi=c\frac{\nu_{o}-\nu_{i}}{\nu_{i}},
\label{eq:doppler}
\eq
where $\nu_{o}$ is the rest frequency of the water transition
(22.23508 GHz), and $\nu_{i}$ is the observed frequency of the
$i^{th}$ maser. Our ability to decompose the model velocities into
classical and relativistic terms (Equation~\ref{eq:dumb11}) is a
convenient by-product of using the optical Doppler shift expression.

The \chisq\ of the model fit is given by:
\bq
\chisq=\sum_{i}\frac{(\vmeasi-\epsilon\vmodi)^{2}}{\sigmavi^2+\beta^2\vroti^2},
\label{eq:chisq}
\eq 
where the summation is over all high-velocity maser features,
\sigmavi\ are the combined uncertainties in the model and observed
velocities, and $\epsilon$ and $\beta$ are geometric correction
factors discussed below.  Because of the excellent spectral resolution
of the VLBA, \sigmavi\ is dominated by uncertainties in the model
velocities as a result of uncertainties in the measured positions of
the masers: 
\bq
\sigmavi\simeq\frac{1}{2}\vroti\frac{\sigxi}{\xxi-\xo},
\label{eq:yuck1}
\eq 
where \sigxi\ is the uncertainty in the right ascension of the
$i^{th}$ feature, given approximately by $\Theta_{x}/(2\mbox{SNR})$.
Here, $\Theta_{x}$ is the x-component of the synthesized beam and SNR
is the signal-to-noise ratio.  We are justified in ignoring the
contribution of \sigyi\ to \sigmavi\ because of the east-west
orientation of the disk. $\epsilon$ and $\beta$ in
Equation~\ref{eq:chisq} correct for the oversimplification of assuming
that all the masers lie precisely along the midline.  Observations of
high-velocity maser accelerations indicate a $1\sigma$ scatter about
the midline of $4.9^{\circ}$ (Bragg \etal\ 2000).  In the present
analysis, we assume that the masers are scattered with a uniform
probability distribution about the midline by an angle \sigmath, taken
to be independent of radius. In Equation \ref{eq:chisq}, $\epsilon$
accounts for the fact that this scatter will impose a downward bias on
the observed LOS velocities. Similarly, $\beta$ specifies the
magnitude of the additional uncertainty in the velocities as a result
of this scatter.  These terms are related to \sigmath\ as follows: \bq
\epsilon\simeq1-\frac{\sigmath^2}{2}+\frac{\sigmath^4}{8}
\label{eq:cool1}
\eq
and
\bq
\beta\simeq\left[\frac{\sigmath^4}{2}\right]^{1/2}.
\label{eq:cool2}
\eq

The \chisq\ in Equation~\ref{eq:chisq} is minimized, via a downhill
simplex algorithm, by varying \xo, \yo, \vo, \mass, and $i$, where the
disk center is measured with respect to the phase center of the
observations, corresponding to the position of the maser at 510 \kms.
Each line in the high-velocity spectra was associated with a distinct
maser feature, and positions were derived as the flux-weighted average
across each line.  The four ``L''-setup epochs listed in Table~1 provide
a total of 99 red features and 17 blue features for the fits.
Ultimately, it is the relative paucity of blue-shifted features that
constrains the quality of the fits.

In practice the high-velocity rotation curve can only be used to
constrain three of the disk parameters, since $\mass^{1/2}$ and $\sin
i$ are degenerate, and since the approximate east-west orientation of
the disk renders the high-velocity fit insensitive to \yo.  The
results of this section are insensitive to \yo, and we use $\yo=0.55$
mas as estimated by Herrnstein (1997), based on high-velocity maser
spatial symmetry considerations.  The inclination can be determined
from the positions and velocities of the systemic masers.  Herrnstein
(1997) reports $i=82^{\circ}\pm1$ at a radius of 3.9 mas for a global
model incorporating a radius-dependent position angle, and a constant
inclination.  Here, we adopt $i=81.4^{\circ}$ at 3.9 mas because this
results in a consistency between the positions of the systemic masers
and the bowl model derived from the high-velocity data.  In fact, the
conclusions of this section are not substantively affected by the
choice of inclination over a range of values consistent with
Herrnstein (1997).

\begin{figure}[p]          
\centering
\includegraphics[width=3in]{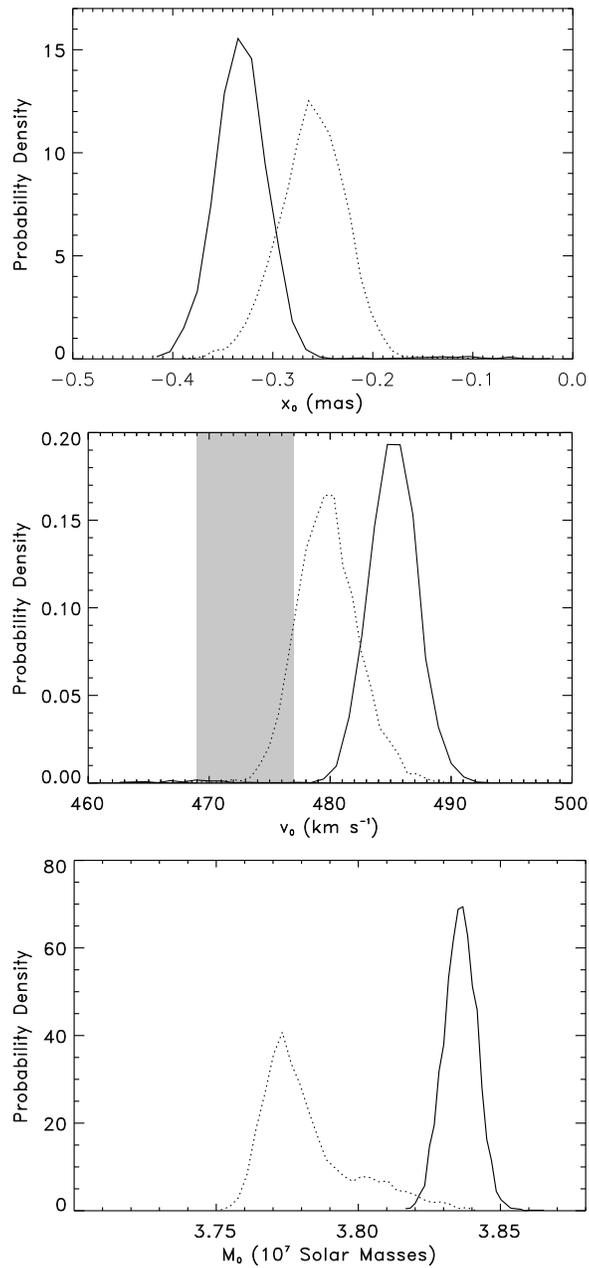}
\caption {Probability density functions for each of the disk center
parameters.  The solid lines assume a pure Keplerian rotation in a
constant inclination disk, while the dashed lines are derived for a
model incorporating an inclination warp.  The shaded region in the
middle plot shows the $1\sigma$ confidence interval for the galaxy 
systemic velocity, as measured from optical observations.}
\label{fg:pdfs}
\end{figure}

\begin{figure}[p]
\epsscale{0.9}          
\plotone{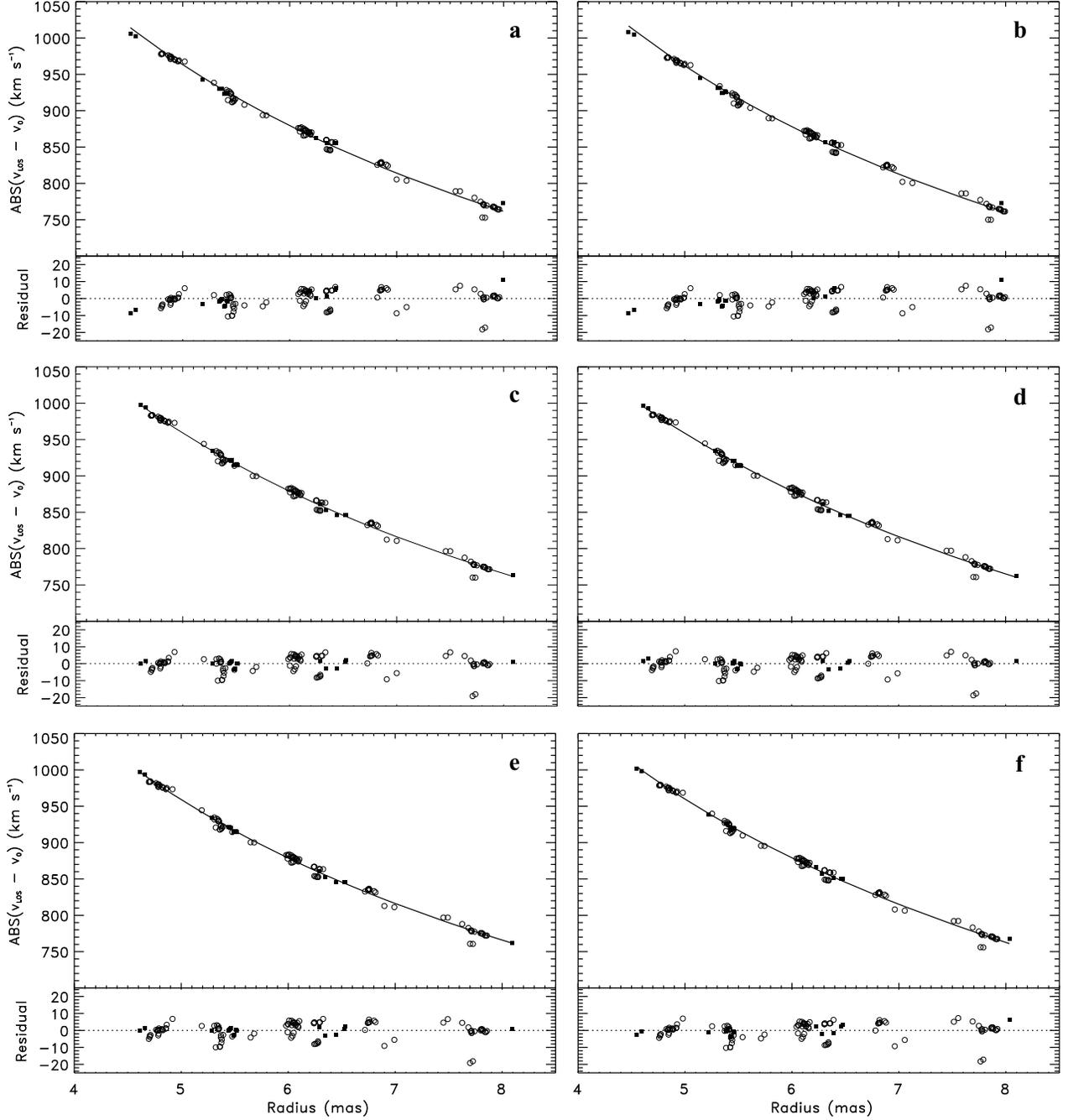}
\caption {Best-fitting high-velocity rotation curve models derived
using epochs 1--4 in Table~1 (solid lines). Open circles mark
positions and velocities of red maser features, filled squares of the
blue features. Residuals are shown in the lower panels.  Panel {\it a}
shows a Keplerian model, neglecting relativistic effects.  All
subsequent panels incorporate relativity. {\it b)} Keplerian
model.  {\it c)} Power law model ($\vrot\propto r^{-\xi})$. {\it
d)} Central cluster model. {\it e)} Massive disk model. {\it f)}
Inclination warp model.}
\label{fg:rcurves}
\end{figure}

The results of our analyses are shown in Table~2.
The first two rows of Table~2 show the disk center parameters for the
best-fitting pure-Keplerian rotation curve model, neglecting and
incorporating relativistic effects, respectively.  The quoted
uncertainties are based on the probability density functions of
Figure~\ref{fg:pdfs}, which are derived via a Monte Carlo analysis in
which \chisq\ minimizations are performed on an ensemble of idealized
datasets with sampling and noise commensurate with the true data.
Panels $a$ and $b$ of Figure~\ref{fg:rcurves} show rotation curves and
residuals for these models.  A \sigmath\ of $4.8^{\circ}$,
corresponding to a velocity scatter of $\sim4.5$ \kms\ at $900$ \kms,
is needed to bring the reduced \chisq\ close to one.  Models
positioning the masers precisely along the midline yield reduced
\chisq\ of over 200. Hence, the rotation curve fits provide a highly
significant statistical detection of the scatter around the midline
that is in good agreement with the Bragg \etal\ (2000) measured value
of $4.9^{\circ}$.

As reported in earlier results, the projected rotation curve is at
least qualitatively well-fit by a Keplerian rotation law.  The data do
not yield an unambiguous detection of relativistic effects primarily
as a result of parameter correlation and the relatively sparse blue
coverage.  These effects are nevertheless incorporated in all
subsequent models.

\subsection{Refinements \label{sec:refine}}

We search for deviations from pure Keplerian motion by considering
four different modifications to the central point-mass, Keplerian
rotation, constant-inclination disk model discussed above.

\subsubsection{Model 0 -- Non-Keplerian Rotation}

\begin{figure}          
\centering
\includegraphics[width=3.5in, angle=270]{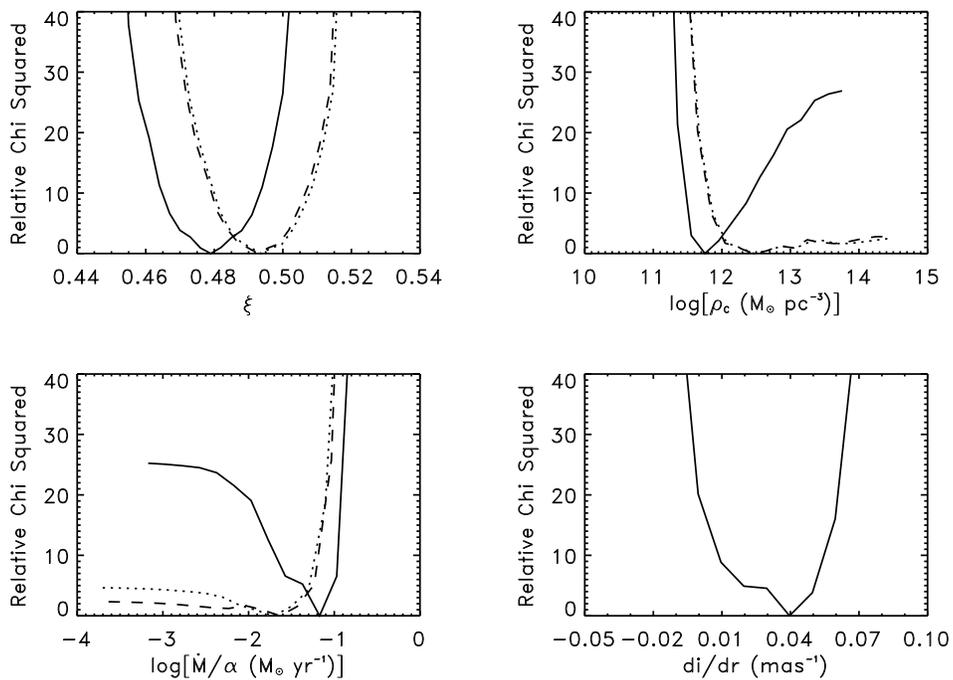}
\caption {Relative \chisq\ for four different refinements to a Keplerian,
constant-inclination disk.  Clockwise from upper left they are: 1)
Power law rotation parameterized as $v\propto r^{-\xi}$; 2) a
central cluster with core density \rhoc; 3) a massive disk,
parameterized in terms of the accretion rate \mdot; and 4) a disk with
a radial gradient in inclination of \didr.  The solid curves show the
effect of these perturbations on a constant-inclination disk, 
the dashed and dotted curves on a disk incorporating 
$\didr=0.04$\,\rpm\ and $\didr=0.034$\,\rpm, respectively.}
\label{fg:chisq}
\end{figure}

\begin{figure}
\plotone{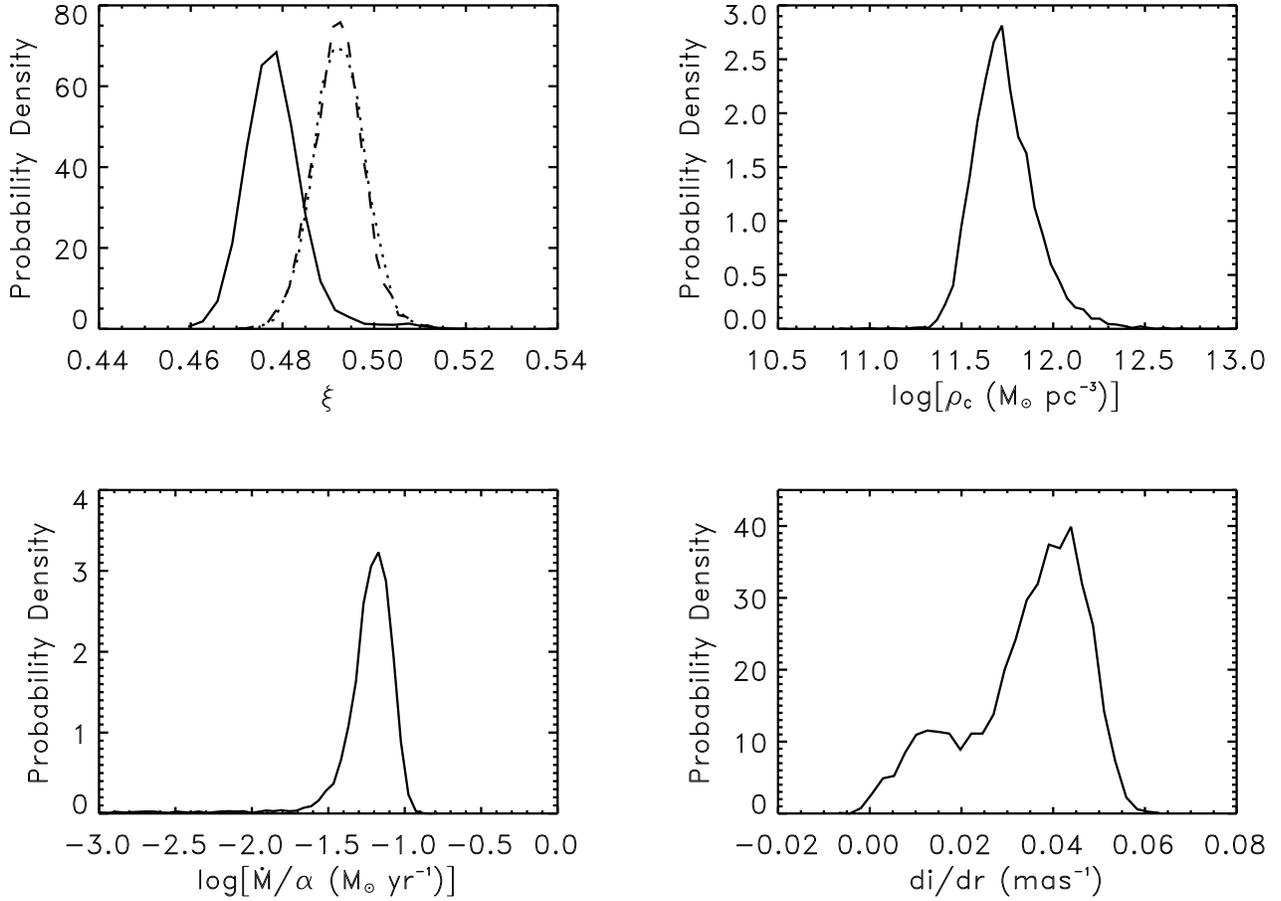}
\caption {Probability density functions for each of the rotation curve
perturbations considered in Figure~\ref{fg:chisq}.  For the power law model
(upper-left panel), the solid curve shows the distribution of $\xi$
for a constant inclination disk, while the dashed and dotted curves show 
the distribution for a disk incorporating $\didr=0.04$\,\rpm\ and 
$\didr=0.034$\,\rpm, respectively.}
\label{fg:pdfs_pert}
\end{figure}

This first refinement is the most direct and arguably the least
enlightening probe of the NGC~4258 rotation curve.  We
search for non-Keplerian rotation across the high-velocity masers by
replacing Equation~\ref{eq:kepler} by 
\bq
\vroti=\frac{\sqrt{G\mass}}{\left[(\xxi-\xo)^{2}+(\yi-\yo)^{2}\right]^{\xi}},
\label{eq:power}
\eq 
and including $\xi$ as a fourth parameter in the maximum likelihood
analysis.  The discrepancy between $\xi$ and 0.5 measures the
deviation from Keplerian rotation in the LOS velocities of the
masers.  Table~2 and panel $c$ of Figure~\ref{fg:rcurves} show the
results of such an analysis. The $10\%$ reduction in the reduced
\chisq\ is highly significant, and the systematics of the blue
residuals are noticeably improved.  Figure~\ref{fg:chisq} shows
relative \chisq\ as a function of $\xi$, and Figure~\ref{fg:pdfs_pert}
shows the $\xi$ probability density function, derived from the Monte
Carlo technique described above.  {\it We find $\sim2\sigma$ evidence
for deviation from Keplerian rotation ($\xi\neq 0.5$) in the
NGC~4258 disk, with a preferred rotation law of $r^{-0.48\pm0.01}$}.
This amounts to a total flattening in the high-velocity rotation curve
of about $9.0$\,\kms\ across the high-velocity masers.

\subsubsection{Model 1: -- Central Cluster}

Maoz (1995) explored the possibility that the binding mass at the
center of the NGC 4258 disk is not a supermassive black hole, but a
spherically symmetric cluster of objects.  Since such a cluster cannot
have abrupt edges, Maoz was able to place a lower limit on the cluster
core density based on the precision of the Keplerian rotation
curve. However, this analysis relied on the relatively crude upper
limits on the deviation from Keplerian rotation available at the time,
and we duplicate it here using the \chisq\ formalism developed in the
previous section.

Following Maoz, we consider the possibility that all the binding mass
resides in a spherically symmetric Plummer-model cluster at the center
of the disk.  The mass distribution is then parameterized in terms of
the core density and radius (\rhoc\ and \rc) as follows: 
\bq
M(<r)=\frac{4\pi\rhoc}{3}r^{3}\left(1+\frac{r^{2}}{\rc^{2}}\right)^{-3/2}.
\label{eq:mass}
\eq 
The cluster model is implemented in the \chisq\ analysis by
associating $M(<r_{i})$ with \mass\ in Equation~\ref{eq:kepler} and
adding $\rhoc$ as a fourth parameter in the fit.  Once again, Table~2
indicates a significant improvement in the reduced \chisq\ over pure
Keplerian models, and Figure~\ref{fg:rcurves} shows the fit and its
residuals.

A disk obeying perfect Keplerian rotation would not possess a
$\chisq(\rhoc)$ minimum.  Instead, the relative \chisq\ would
asymptotically approach zero for large values of \rhoc\ and such an
analysis would provide only a lower limit on \rhoc. The minimum in
relative $\chisq(\rhoc)$ evident in Figure~\ref{fg:chisq} therefore
implies a significantly flattened rotation curve. Specifically, we
find $\rhoc=5.7\pm1.4\times10^{11}$\,\smpc (Figure~\ref{fg:pdfs_pert}), a
core radius of $0.78$\,mas ($=8.4\times10^{16}$\,cm), a total cluster
mass between the inner and outermost masers of
$1.1\times10^{6}$\,\msun, and a resulting rotation curve flattening of
10.9 \kms\ across the high-velocity masers.

\subsubsection{Model 2 -- Massive Disk}

We now consider the possibility that mass in the outer accretion disk
itself is responsible for flattening the rotation curve.  The maser
observations provide direct evidence that the outer disk is thin,
cool, and dense. For a thin, isothermal, steady-state accretion disk
in hydrostatic equilibrium (c.f. Frank, King, \& Raine 2002), the
scale height is $\cs r^{3/2}(GM)^{1/2}$, and the midplane density as a
function of radial distance $r$ can be written (Neufeld \& Maloney
1995): \bq \rhomid=\frac{G\mo\mdot}{3\pi(2\pi)^{1/2}\alpha
r^{3}\cs^{3}}
\label{eq:centraldensity}
\eq 
where \mo\ is the central mass, \mdot\ is the accretion rate, \sound\
is the sound speed, and $\alpha$ is the 
Shakura-Sunyaev parameter of the kinematic viscosity.
Integration of Equation~\ref{eq:centraldensity} gives the mass of disk, \dmass,
 within the inner
and outer boundaries of the maser emission, $r_1$ and $r_2$,         
\bq
\dmass=8.3\times10^{4}\frac{\mdot}{\alpha}\frac{\mo^{1/2}\left(r_{2}^{1/2}-r_{1}^{1/2}\right)}{\sound^{2}}\mbox{
\msun},
\label{eq:ss1}
\eq 
where  $r_{1}$ and $r_{2}$ are in
pc, \sound\ is in \kms, \mdot\ is in solar masses per year and \mo\ is in solar masses. 
 The disk
mass internal to the $i^{th}$ maser feature (\dmassi) is estimated
from Equation~\ref{eq:ss1}, using $r_{2}=\ri$ and $r_{1}=\rmin$ (the
radius of the innermost maser), and assuming $\sound=2.3$\,\kms,
corresponding to a temperature of 800\,K. Theoretical models predict
that temperatures between 500 and 1000\,K are most conducive to the
formation of water masers.

A massive disk is included in the rotation curve model by
incorporating \mdota\ as a fourth parameter in the fit and replacing
\mass\ in Equation~\ref{eq:kepler} by $\mass+\dmassi/d$, where the
distance, $d$, is taken to be 7.2\,Mpc.  The reduced \chisq\
improvement is comparable to the cluster model results (Table~2), and
the rotation curve and residuals are included in
Figure~\ref{fg:rcurves}.  As with the central cluster analysis of the
previous section, the pronounced minimum in relative $\chisq(\mdota)$
evident in Figure~\ref{fg:chisq} implies a significantly flattened
rotation curve. The \chisq\ analysis favors an accretion rate of
$6.5\pm1.3\times10^{-2}\alpha$\,\smy (Figure~\ref{fg:pdfs_pert}),
which leads to a total disk mass of $8.9\times10^{5}$\,\msun\ and a
$\sim10.1$\,\kms\ flattening in the rotation curve across the
high-velocity masers. Note that for this disk mass the midplane
densities at the inner and outer radii of 0.14 and 0.28 pc are $4
\times 10^{11}$ and $0.8 \times 10^{11}$ \percmcmcm\, respectively,
about an order of magnitude above the normally preferred range for
maser emission of $10^{8}$ to $10^{10}$ \percmcmcm. Hence, the disk
conditions implied by the existence of maser emission make this model
unlikely.


\subsubsection{Model 3 -- Inclination Warp \label{sec:fit}}

The disk geometry is parameterized in terms of two Euler angles: the
position angle (\pang) is azimuthal about the LOS while the
inclination angle is polar. A warped disk can be described by allowing
these angles to vary with radius.  While the specific functional
relationship between $i$ and \pang\ and $r$ will depend on the
perspective of the observer, it can be expected that both
Euler angles will have some radial dependence. As discussed in
Section~1, the declinations of the high-velocity masers are well-fit
by a position angle obeying a second order polynomial in radius.
Thus, it seems likely that the inclination of the disk will also vary
with radius.  Such an inclination warp will manifest itself as an
apparent deviation from pure Keplerian rotation in the LOS velocities 
of the high-velocity masers.

We quantify this effect by postulating the following ({\it ad hoc}) linear 
inclination warp model:
\bq
\ii=\io+\frac{di}{dr}(\ri-\ro),
\label{eq:iwarp}
\eq 
where \ii\ is the inclination of the disk at the radius of the
$i^{th}$ maser. As discussed above, \io\ and \ro\ are fixed at
$81.4^{\circ}$ and 3.9~ mas, respectively, based on observations of
the systemic masers.  The inclination warp is added to the projected
rotation curve fit by replacing $i$ by \ii\ in Equation~\ref{eq:dumb1}
and including \didr\ as a fourth free parameter.  This leads to a
significant improvement in the reduced \chisq\ over pure Keplerian,
constant-inclination models (Table~2) and the model and residuals are
shown in Figure~\ref{fg:rcurves}.  The \chisq\ analysis prefers an
inclination gradient of $0.04\pm0.01$\,\rpm\ across the high-velocity
masers (Figures~\ref{fg:chisq} and \ref{fg:pdfs_pert}).  Consistent
with previous results, this corresponds to a rotation curve flattening
of about 9.6\,\kms\ across the high-velocity emission. In this model,
the inclination of the disk changes from $81.4^{\circ}$ at the radii
of the systemic masers (3.9 mas), to about $91^{\circ}$ at the radius
of the outermost high-velocity feature (8.1 mas). The magnitude of the
inclination warp is approximately equivalent to that of the position
angle warp.

We emphasize that, unlike the other models under consideration, the
inclination warp model does not indicate non-Keplerian motions in the
NGC~4258 disk, but rather radius dependent projection effects in
the disk.

\subsection{Discussion \label{sec:disclaimer}}

\begin{figure}[bht]          
\begin{center}
\includegraphics[width=3.5in, angle=90]{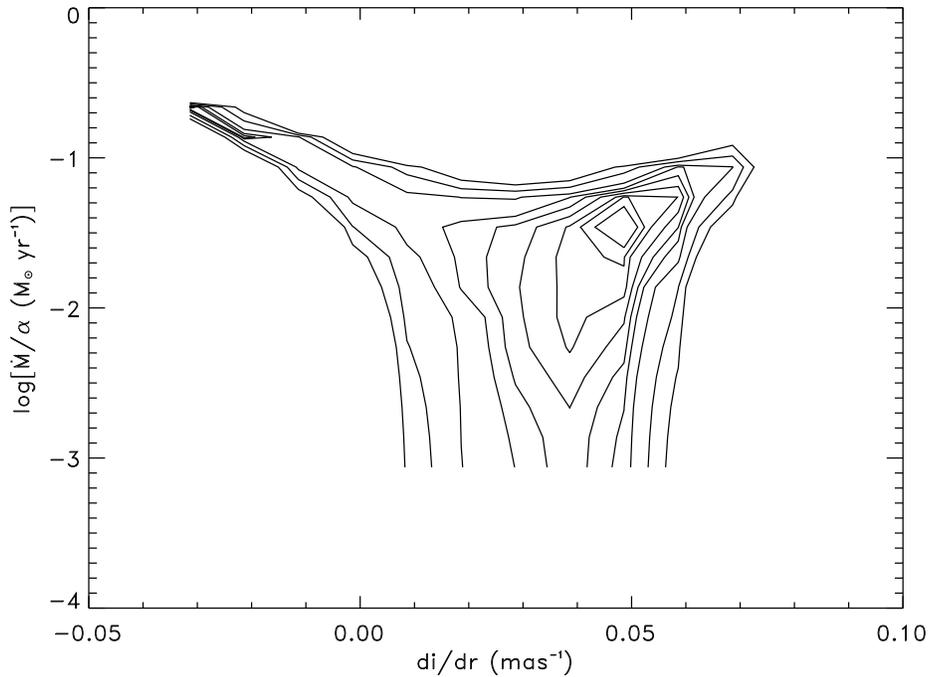}
\end{center}
\caption {Simultaneous relative \chisq\ intervals for \mdot\ and
\didr.  Contour steps are 1, 1.5, 2, 3, 4, 5, 7, 11, and 15.  The
strong correlation between the effects of a massive disk and an
inclination warp on the rotation curve is evident.}
\label{fg:joint}
\end{figure}

While there is compelling evidence for a $\sim8$\,\kms\ flattening in
the projected rotation curve of the high-velocity masers, its physical
origin is ambiguous.  In the previous sections, we have examined three
possible mechanisms, each in principle capable of generating the
observed rotation curve. Not surprisingly, the parameters defining
these three mechanisms are highly correlated with one another.
Furthermore, because an inclination warp is capable of both flattening
and steepening the rotation curve, the actual disk could in fact
incorporate any combination of these effects. This is demonstrated
graphically in Figure~\ref{fg:joint}, which shows contours of constant
\chisq\ throughout the \didr--\mdota\ plane. The correlation between
the parameters is clear, and arbitrarily large \mdota\ can be
supported by positing negative values for \didr\ without incurring too
dramatic an increase in \chisq.  This limits our ability to place firm
limits on either \mdota\ or \rhoc\ from the high-velocity rotation
curve alone.  We note in passing that the overall minimum in the joint
\chisq\ of Figure~\ref{fg:joint} suggests a detection of the effects
of finite mass in an inclination-warped disk.  However, the detection
is very weak (a $\Delta\chisq$ of $\sim2$ for the inclusion of \mdot\
as a new parameter) and the densities implied by the preferred
accretion rate are inconsistent with the existence of masers.

The three mechanisms considered above are not intended to be an
exhaustive list of the mechanisms capable of flattening the projected
rotation curve. For example, it might be possible to fit the rotation
curve without violating the acceleration data by systematically
displacing the innermost masers farther from the midline than the
outer masers.  We merely note in passing that such a geometry is
counterintuitive in that the envelope of maser emission is expected to
grow with radius, along with the coherent velocity gain paths.
 
\section{The Systemic Masers and the Inclination Warp}

The previous section suggests that additional information is required 
to discern the true origin for the flattened rotation curve. In this 
section we show that the inclination warp derived from the high-velocity 
masers provides a compelling explanation for the morphology of the 
systemic masers. Based on this we conclude that inclination warping is 
indeed the source of the apparently flattened rotation curve.  We begin 
by first demonstrating that there are substantial unanswered questions 
surrounding the geometry of the systemic features.  We then show that 
the inclination warp answers these questions convincingly.

\subsection{The Systemic Masers}

\begin{figure}[bht]          
\centering
\includegraphics[width=3.5in, angle=90]{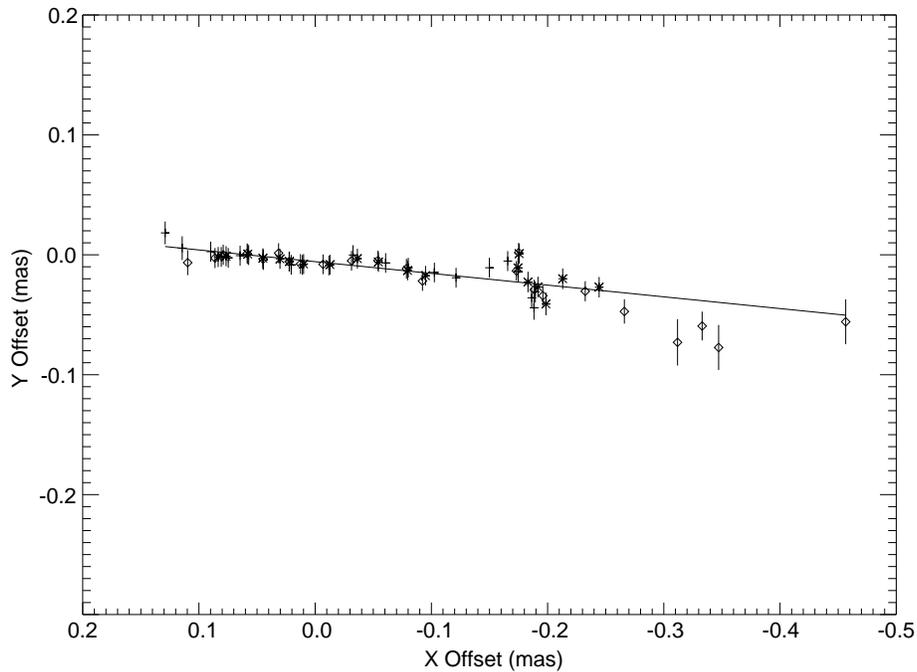}
\caption {Relative RA versus declination for the systemic masers for
epochs BH25a (+), BH25b (*), and BH25c (diamonds).  The best-fitting
line corresponds to a position angle of $5.6^{\circ}$. The error bars
include a non-instrumental noise of 8~\muas\ added in quadrature to
achieve a reduced \chisq\ of 1 in the linear fit.}
\label{fg:xy}
\end{figure}

The initial VLBA observations of the NGC 4258 masers (1994 April)
failed to detect any vertical structure in the systemic masers at the
10\,\muas\ level, leading Moran \etal\ (1995) to conclude that the
disk aspect ratio ($H/r$) must be less than about 0.25\%. We have
repeated this analysis using epochs 6--8 in Table~1. These
observations include the Effelsberg 100-m antenna and are about three
times more sensitive to vertical structure.  The results are shown in
Figure~\ref{fg:xy}.  The slope of the best-fitting line suggests the
systemic masers trace a disk with a $5.6^{\circ}$ position angle.  The
linear fit has a reduced \chisq\ of 82 when the $y$-coordinate
instrumental uncertainties are used, indicating a highly significant
detection of vertical structure in the systemic masers. Specifically,
a reduced \chisq\ of one is achieved by adding in quadrature an
additional scatter in $y$ (\sigyp) of 8\,\muas\ or 0.00024\,pc,
indicating that $H/r\la0.2$\% at the radius of the systemic masers.

\begin{figure}[p]          

\centering
\includegraphics[width=5.5in]{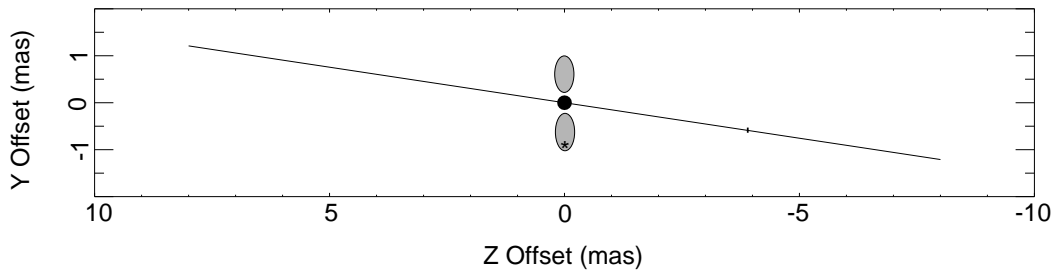}
\caption {Cross-section of a constant-inclination disk. The LOS is
along the $-z$ direction. The northern shaded oval marks the observed
location of northern jet emission, and the southern oval is the
reflection of this through the disk center.  The asterisk marks the
observed location of the southern jet emission.  The small vertical
dash at a Y Offset of $-0.55$ denotes the position of the systemic
masers in the disk, drawn to scale in Y. It is difficult to reconcile
the narrow Y confinement of the systemic masers with background
amplification models and the large radial spread of the high-velocity
masers.}
\label{fg:cross_flat}
\end{figure}

For a constant inclination disk, the narrow spread in $y$ of the
systemic masers is puzzling in light of the fact that the
high-velocity masers extend from 4.7 to 8.1 mas.  It is even more
surprising when we consider that the systemic masers apparently
amplify background continuum emission from the compact nuclear jet.
Figure~\ref{fg:cross_flat} shows, for a constant-inclination disk, $z$
(distance along the LOS) versus $y$ along the radial perpendicular to
the disk midline, with disk azimuth (\thetad) of
$270^{\circ}$. \thetad\ is measured in the direction of rotation from
the ascending node of the disk, such that $\thetad=0$ for the
red-shifted masers.  The LOS velocities of the systemic masers range
from about $\vo-70$\,\kms\ to $\vo+70$\,\kms\ indicating that they
occupy a $\sim7^{\circ}$ arc centered about $\thetad=270^{\circ}$.
This is a natural place for masers to occur since, together with the
midline, it is the location in the disk where $\partial\vlos/\partial
z$ is a minimum.  Furthermore, as long as the dependence of position
angle on radius is not too steep at small radii, jet emission aligned
with the rotation axis of the disk lies behind the
$\thetad=270^{\circ}$ radial irrespective of position angle, and
masers located here are well-situated to amplify background jet
emission.

The vertical dash in Figure~\ref{fg:cross_flat} marks the position and
$y$ spread of the systemic masers.  Continuum emission has been
detected in seven of eight VLBA epochs.  The northern jet emission is
variable in both strength, ranging from less than 2\,mJy to 6\,mJy,
and in position, with northern offsets between 0.35 and 1.0\,mas.  The
southern jet has remained relatively stable at 0.5\,mJy and 1.0\,mas
South of the disk center (Herrnstein \etal\ 1997, 1998a). The observed
range of northern jet emission is marked by the upper shaded oval in
Figure~\ref{fg:cross_flat}.  The southern shaded oval is the
reflection of the northern emission through the disk center, and the
asterisk marks the location of the observed southern
emission. Herrnstein, Greenhill, \& Moran (1996) demonstrate that any
portion of the disk directly exposed to the central X-rays will
develop a thin layer of ionized gas with 22-GHz free-free optical
depths of $\ga2$ that scales as $\sim r^{-2}$. They further
propose that the southern jet appears weaker and is displaced
southward as a result of thermal absorption in the disk.
Figure~\ref{fg:cross_flat} shows that if the jet is symmetric, the
brightest 22 GHz emission lies directly behind the systemic masers
suggesting that these masers amplify southern jet emission that is
otherwise obscured from view by the disk itself. In this scenario, the
extremely bright systemic masers are only mildly saturated, and their
variability is naturally explained by variability in the background
continuum. Both the northern jet continuum and the systemic masers
vary on timescales of weeks.

Clearly there is a serious inconsistency between the background
amplification models and the observed narrow $y$ extent of the
systemic masers at $\thetad\simeq270^{\circ}$.  While the background
22 GHz continuum radiation extends to $-1.0$\,mas with respect to the
disk center, the systemic maser emission is narrowly confined around
$y=-0.55$\,mas.  Furthermore, symmetry with the northern jet emission
would imply that amplifying systemic masers should range in $y$ from
-0.35 mas to -1.0, about a factor of ten larger than the observed
spread.  Of course, the bright systemic masers require {\it both}
background seed photons {\it and} an inverted population of water
molecules. However, it is very difficult to construct a pumping
mechanism that leads to such a narrow annulus of inverted molecules.
Neufeld \& Maloney (1995) show that the high-velocity masers are
pumped by direct exposure to the central X-ray emission, as a result
of the position angle warp.  No plausible inclination warp would yield
such a narrow cusp of pumped water on the near edge of the disk.  One
possibility is that the outer cool disk is truncated within 3.9 mas,
and that the systemic masers trace the narrow annulus of pumped gas
along the inner edge of the disk. However, we know of no physical
explanation for such a truncation.

The narrow spread in the systemic declinations can also place more
quantitative restrictions on a constant-inclination disk.  For a disk
with no inclination warp, the observed scatter in $y$ places a firm
upper limit on the radial scatter (\sigr) of the systemic masers.  For
a disk tipped down $8^{\circ}$ from edge-on, $\sigyp\simeq0.14\sigr$,
and the detected vertical scatter in the systemic masers requires
$\sigr\la0.06$\,mas ($=0.002$\,pc).  Thus, for a constant-inclination
disk, the narrow spread in $y$ of the systemic masers requires a
fractional radial spread of less than 1.5\% at a radius of 3.9\,mas.
This requirement can be tested directly using the velocities of the
systemic masers.

If the systemic masers are indeed distributed azimuthally in a
Keplerian disk, then 
\bq
\frac{d\vlos}{dx}\simeq\frac{G\mass^{1/2}\sin(i)\cos(\phi)}{\rs^{3/2}}=\mbox{
constant}.
\label{eq:da}
\eq
where $\phi$ is the disk position angle.  The instrumental uncertainties
in the velocity are negligible, and we express the uncertainty in $v$,
\sigv, as the quadrature sum of two terms: 
\bq 
\sigv=270 \sqrt{\sigx^2 + \left(\frac{3}{2} \frac{\sigr}{r} \left(x-\xop\right)\right)^2}
\eq
where \sigx\ is the instrumental uncertainty in right ascension and
$\xop=\xo+\ro \sin(\phi) \cos(i_0)$ is the projected near edge of the
disk where the observed LOS velocity is precisely the systemic
velocity and is independent of radius. 270\,\kms\,mas$^{-1}$ is the
well-constrained prior on the position-velocity slope based on
numerous, past observations.  The first term in the sum is the
straightforward propagation of position uncertainties through to
\sigv, while the second propagates \sigr\ through to \sigv.  

The fit in the upper panels of Figure~\ref{fg:xv} uses $\sigr/r=0.015$
in conformation with the upper limit on the fractional scatter in
radius from the observed scatter in $y$.  The resulting reduced
\chisq\ of 40 demonstrates a significant inconsistency in the
distributions of the declinations and velocities of the systemic
masers in the context of a constant-inclination disk.  In the lower
panels a reduced \chisq\ of one is achieved for $\sigr/r=0.13$, a
factor of 10 larger than indicated by the declination data.  Hence,
{\it a constant-inclination disk model is unable to accommodate both
the extremely narrow scatter in declination and the more substantial
velocity deviations of the systemic masers.}

\subsection{The Inclination Warp: Masers in a Bowl \label{sec:bowl}} 

In the rotation curve analysis of Section~\ref{sec:fit} we found
$\didr=0.04\pm0.01$\,\rpm.  In the sections that follow, we will
assess this estimate based on a variety of considerations involving
the systemic masers.  The resulting refined estimate for the
inclination warp, \didrp, of 0.034\,\rpm\ is the best-guess estimate
for the warp based on the high-velocity and systemic data together.
In motivating the inclination warp in the present section, we are
justified in using this value for the warp, as it is well within the
uncertainties indicated by the high velocity data alone
(Figure~\ref{fg:pdfs_pert}).

\begin{figure}[hp]          
\centering
\includegraphics[width=5in]{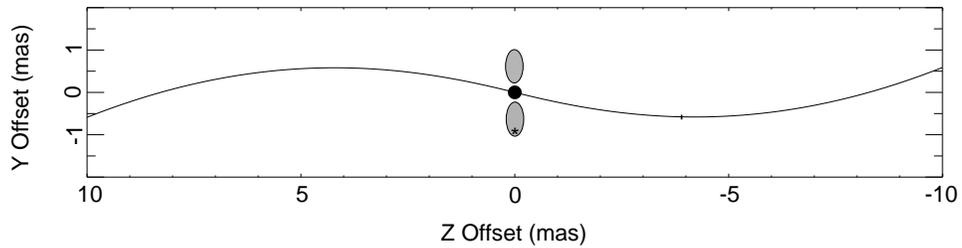}
\caption{Cross-section of an inclination-warped disk with
$\didr=0.034$\,\rpm. The inclination angle measured downward from the
LOS is $i=0.28+0.34z$, where $z$ is in milliarcseconds and $i$ is in
radians, and the shape of the warp is $y=z\sin{i}$.  The LOS is along
the $-z$ direction. The northern shaded oval marks the observed
location of northern jet emission, and the southern oval is the
reflection of this through the disk center.  The asterisk marks the
observed location of the southern jet emission. The small vertical
dash at a Y offset of $-0.55$\,mas denotes the position of the
systemic masers in the disk, drawn to scale in Y. }
\label{fg:cross_bowl}
\end{figure}

Figure~\ref{fg:cross_bowl} shows the $\thetad=270^{\circ}$
cross-section of a disk with $\didr=0.034$\,\rpm.  At the observed
$y$-offset of the systemic masers, marked by the vertical dash in the
figure, $\partial y/\partial r\vert_{(\theta=270)}\approx 0$. In other
words, for a \didr\ consistent with the high-velocity rotation curve,
the systemic masers lie at the bottom of a bowl along the near edge of
the disk, as viewed from our perspective. That the systemic masers lie
on the disk at all at $y=-0.55$\, is critically sensitive to our
choices for \yo, \io, and \ro, and indeed determined the values used
for these parameters in the fits of Section~\ref{sec:fit}.  However,
the fact that these high-velocity fits lead to a geometry in which the
systemic masers lie at a tangent point to the disk is either a
remarkable coincidence or of physical significance.

\begin{figure}[hb]          
\centering
\includegraphics[width=5in]{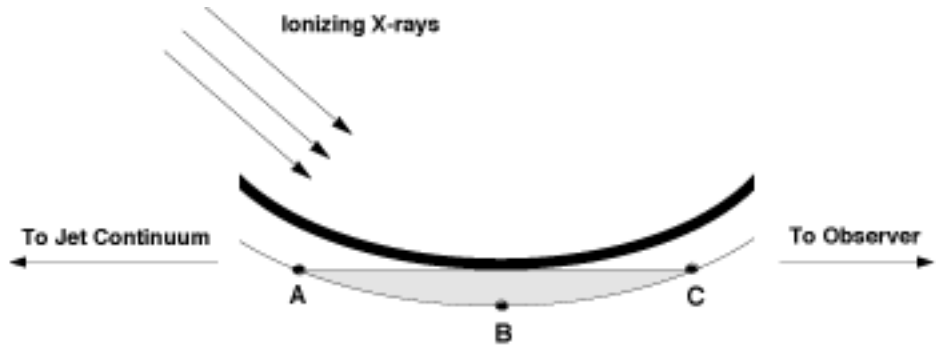}
\caption{Schematic drawing of the bowl bottom.  Both the thickness of
the bowl as well as the curvature have been exaggerated.  The thick
line on the upper surface of the disk represents a thin layer of gas
ionized by x-ray emission from the central engine.  The shaded region
demarcates the expected region of brightest maser emission. See text
for a discussion of points {\it A}, {\it B}, and {\it C}.}
\label{fg:cartoon}
\end{figure}

In fact, there is a compelling reason the brightest systemic emission
{\it should} occur at the bottom of such a
bowl. Figure~\ref{fg:cartoon} is a schematic representation of the
bowl cross-section.  The upper surface of the bowl is directly exposed
to X-rays from the central engine and will therefore possess the
aforementioned thin surface layer of ionized gas with optical depths
$\ga2$. Any molecular gas suitable for supporting masers will lie
underneath this ionized gas. Three radii are labeled in the Figure:
{\bf B} marks the bottom of the bowl, while points {\bf A} and {\bf C}
delineate the range of radii with unattenuated sightlines to both the
background continuum and us.  Water molecules internal to {\bf A} will
have an unobstructed view of the background jet. However, any maser
emission generated in this gas must traverse two ionized surfaces of
the bowl along the LOS to us and will therefore be attenuated by a
factor $\ga50$. Similarly, while molecular gas external to {\bf C}
will have an unobstructed LOS to us, it will have an equally
attenuated view of the background continuum and will effectively be
limited to self amplification.  By analogy with the self-amplified
high-velocity masers, such emission is likely to be at least 20 times
weaker than the background amplified emission in the shaded region at
the bottom of the bowl.  

A significant shortcoming of the constant-inclination disk model is
its inability to explain the narrow $y$-confinement of the systemic
masers (Figure~\ref{fg:cross_flat}). In contrast, the bowl model
predicts that the brightest systemic emission ought to lie at the bowl
bottom in a narrow declination band on the sky, corresponding to the
thickness of the masing region along the bottom of the bowl (shaded
region in Figure~\ref{fg:cartoon}).  Figure~\ref{fg:cross_bowl} shows
that this is indeed the case for the \didr\ estimated from the
high-velocity data.  In addition to predicting that the brightest
systemic emission arises in the bowl bottom, the bowl model also
allows for much ($\sim20-50$ times) fainter systemic emission at
larger or smaller radii, as described above.  Such emission is not
ruled out in the present analysis.

\begin{figure}[hb]          
\centering
\includegraphics[width=5in]{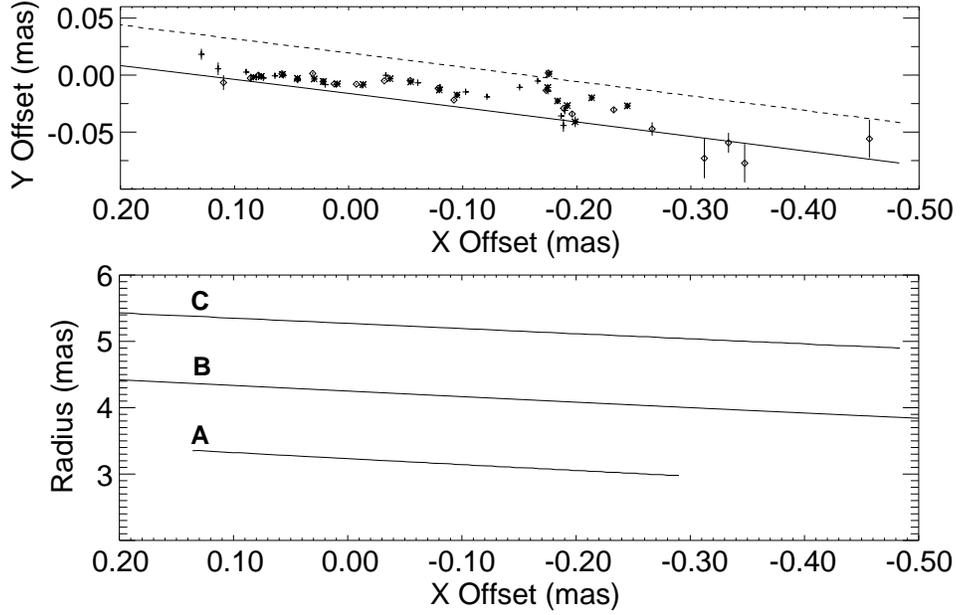}
\caption {{\it Top:} Relative RA versus declination for the systemic
masers for epochs BH25a (+), BH25b (*), and BH25c (diamonds). Error
bars are the instrumental uncertainties.  The solid line marks the
bottom of the bowl for $\didr=0.034$\,\rpm.  The dotted line
corresponds to a $1\sigma$ maser layer thickness of 15~\muas. {\it
Bottom:} Radius as a function of $x$ for the inner edge {\bf A},
bottom {\bf B}, and outer edge {\bf C} of the bowl (as identified in
Figure~\ref{fg:cartoon}) assuming a maser layer thickness of 15\muas.}
\label{fg:xy_bowl}
\end{figure}

The top panel of Figure~\ref{fg:xy_bowl} once again shows systemic
maser $x$ versus $y$. The solid line demarcates the bowl bottom, and
demonstrates in detail the coincidence in declination of the systemic
masers and the bottom of the bowl.  The parallel dotted line is
intended to represent the approximate maser-layer thickness needed to
accommodate all the systemic emission.  It corresponds to a $1\sigma$
thickness of 15\muas.  It is larger than the $1\sigma$ thickness
inferred from a straightforward linear fit to the data
(Figure~\ref{fg:xy}) because the bowl bottom is not parallel to this
best-fitting line.  Figure~\ref{fg:xy_bowl} suggests that the
declinations of the systemic masers require a masing layer of
$\sim15\muas$ ($1\sigma$) at the bottom of the bowl.  The bottom panel
of Figure~\ref{fg:xy_bowl} shows radius as a function of $x$ for the
three points identified in Figure~\ref{fg:cartoon}.  There is a
$\sim10\%$ gradient in the radius of the bowl bottom across the
envelope of systemic emission.  The inner and outer edges of the bowl
({\bf A} and {\bf C}) are located 0.77 and 1.23 times the bowl bottom
radius for a maser layer thickness of 15\muas.

\begin{figure}[bht]          
\centering
\includegraphics[width=5in]{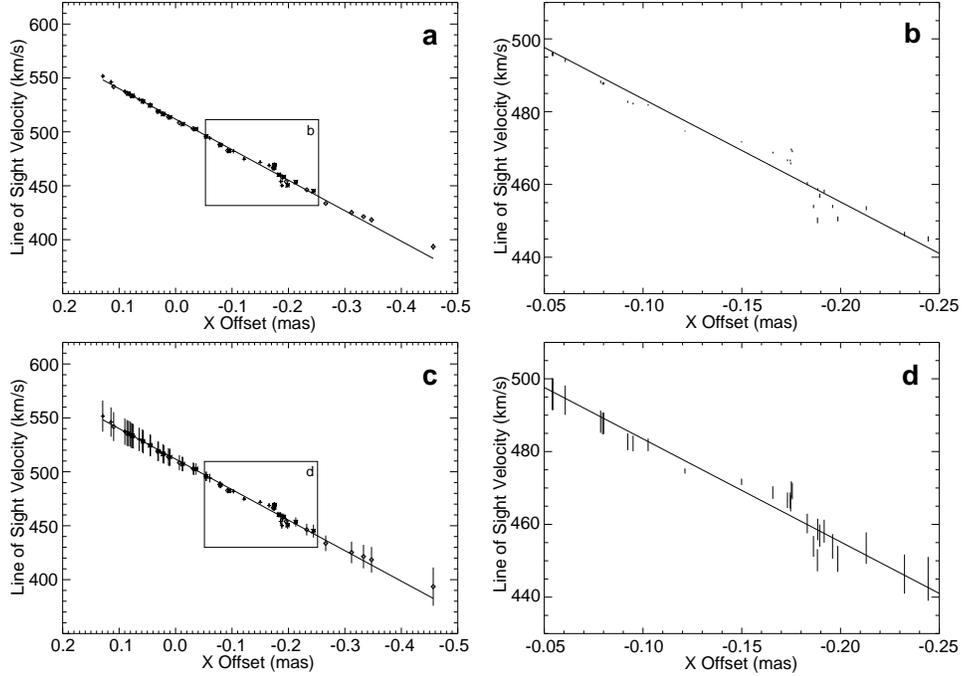}
\caption { Relative RA versus LOS velocity for the systemic masers for
epochs BH25a (+), BH25b (*), and BH25c (diamonds). The fits were
performed assuming $\xo=-0.19$\,mas and $\phi=5.6^{\circ}$ (see text
for details). {\it a)} and {\it b)} Linear fit assuming
$\sigr/r=0.015$, yielding a slope of 270\,\kms\,mas$^{-1}$ and a
reduced \chisq\ of 40.  {\it c)} and {\it d)} Fit using $\sigr/r=0.13$
giving a slope of 266\,\kms\,mas$^{-1}$ and a reduced \chisq\ of 1.
The insets show in greater detail the portion of the disk that
dominates the \chisq.}
\label{fg:xv}
\end{figure}

A second serious deficiency of the constant-inclination disk is an
inconsistency between the extremely narrow declination scatter and
relatively larger velocity spread of the systemic masers
(Figure~\ref{fg:xv}).  Figure~\ref{fg:xy_bowl} shows that masers along
the bowl bottom, and therefore confined to a narrow band of declination
on the sky, can, in principle, reside over a large range of radii in the
disk.  Because the bowl model removes the proportionality between $y$
and $r$ characteristic of constant-inclination models, it also removes
any inconsistency between the declinations and velocities of the
systemic masers.

\begin{figure}[ht]          
\centering
\includegraphics[width=3.5in, angle=90]{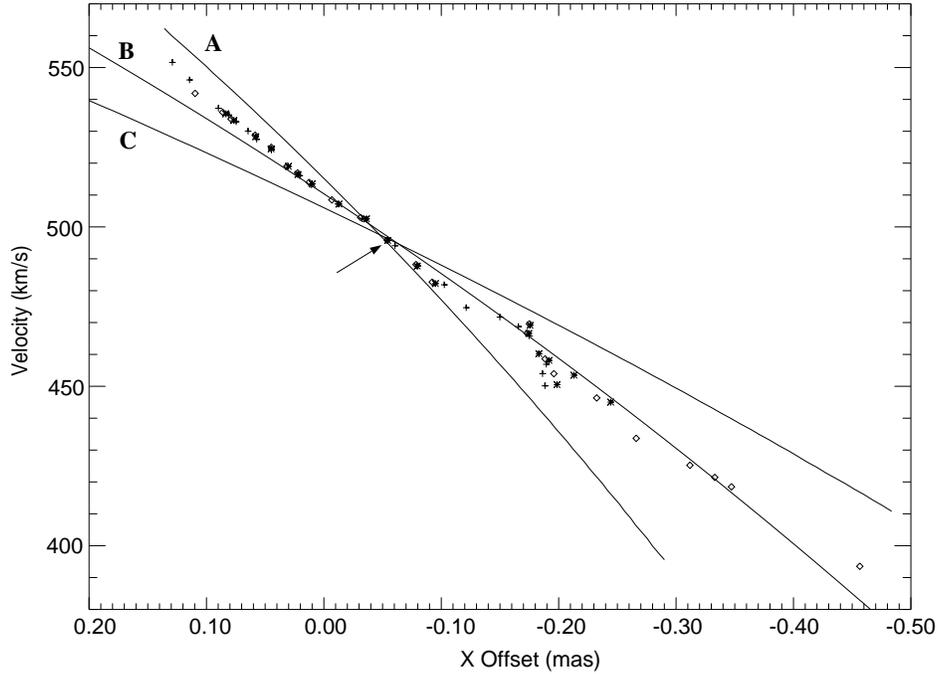}
\caption{LOS velocity versus $x$ offset for the systemic masers for
epochs BH25a (+), BH25b (*), and BH25c (diamonds). The solid lines
show model LOS velocity as a function of $x$ for the positions
indicated in Figure \ref{fg:cartoon}, assuming a maser-layer thickness
of 15~\muas, and using $\vo=473.5$\,\kms\ and $\xo=-0.19$\,mas.  The
arrow indicates the maser features associated with the flare, detected
in each of the three epochs.}
\label{fg:3line}
\end{figure}

This conclusion is quantified in Figure~\ref{fg:3line}, which shows
actual systemic maser position-velocity data superposed on the
predicted LOS velocity for the three annuli identified in
Figure~\ref{fg:cartoon}, assuming a maser layer thickness of 15\muas.
The precise relationship between the model $x$ and $v$ depends on both
the inclination and position angle warps.  As in previous works
(Herrnstein, Greenhill, \& Moran 1996), we estimate the latter warp by
assuming the high-velocity masers lie along the disk midline, and
fitting a second order polynomial through their positions.  Using the
first four epochs of Table~1, we find the data are well fit by
$\phi=0.425-0.088r + 0.0038r^2$ ($r$ in mas), in good agreement with
past estimates for the position angle warp.

If the systemic masers were positioned exactly along the bowl bottom,
we would expect the data to lie along the line {\bf B} in
Figure~\ref{fg:3line}.  However, given the scatter in the systemic
declinations and the bowl curvature determined from the high-velocity
data, the systemic masers could in principle lie anywhere in the range
of radii shown in Figure~\ref{fg:xy_bowl}.  Hence, the lines {\bf A}
and {\bf C} show as a function of $x$ the envelope of LOS velocity
that is consistent with the bowl model and the observed scatter in
$y$.  Whereas the constant inclination disk is unable to handle the
structure in the systemic $x$--$v$ data, and in particular the
structure between $-0.05$ and $-0.25$ mas (Figure~\ref{fg:xv}), the
bowl model has no trouble accommodating the observations.

\begin{figure}[ht]          
\centering
\includegraphics[width=4in]{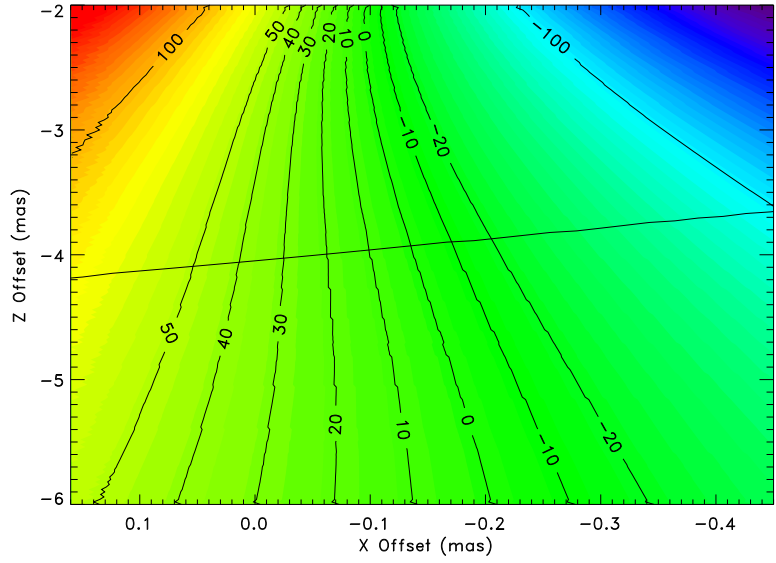}
\caption{Contours of constant LOS velocity relative to
$\vo=473.5$\,\kms along the bottom of the bowl.  The LOS is along the
$-z$ direction, and hence vertical contours indicate a zero velocity
gradient across the bowl along the LOS.  The model used
$\xo=-0.29$\,mas.  The roughly horizontal line marks the position of
the bottom of the bowl (B in Figure \ref{fg:cartoon}).  The
intersection of this line with the zero contour locates the projected
near edge of the disk (\xop) at $\sim-0.15$\,mas.}
\label{fg:isovel}
\end{figure}

The model velocities in Figure~\ref{fg:3line} were generated using
$\vo=473.5$\,\kms\ and $\xo=-0.19$\,mas.  The convergence point of the
three velocity curves in the figure marks the point along the near
edge of the disk where the LOS velocity gradient across the bowl
bottom and along the line of sight reaches a minimum of $\la2$\,\kms.
A noteworthy feature of the model is that the LOS velocity gradient
across the bowl is a minimum for a velocity and a position that are
$\sim20$\,\kms\ and $\sim0.1$\,mas to the red of the systemic velocity
and projected near edge of the disk ($\xop\simeq-0.15$\,mas),
respectively.  This is also evident in Figure~\ref{fg:isovel}, which
shows lines of constant LOS velocity in the bowl assuming
$\xo=-0.19$\,mas. The iso-velocity lines are most nearly vertical for
velocities $\sim20$\,\kms\ red of \vo.  In the next section we make an
association between this velocity convergence point and the observed
systemic flare.
	
\subsection{The Inclination Warp: The Systemic Flare \label{sec:flare}} 

\begin{figure}[ht]          
\centering
\includegraphics[width=3.5in]{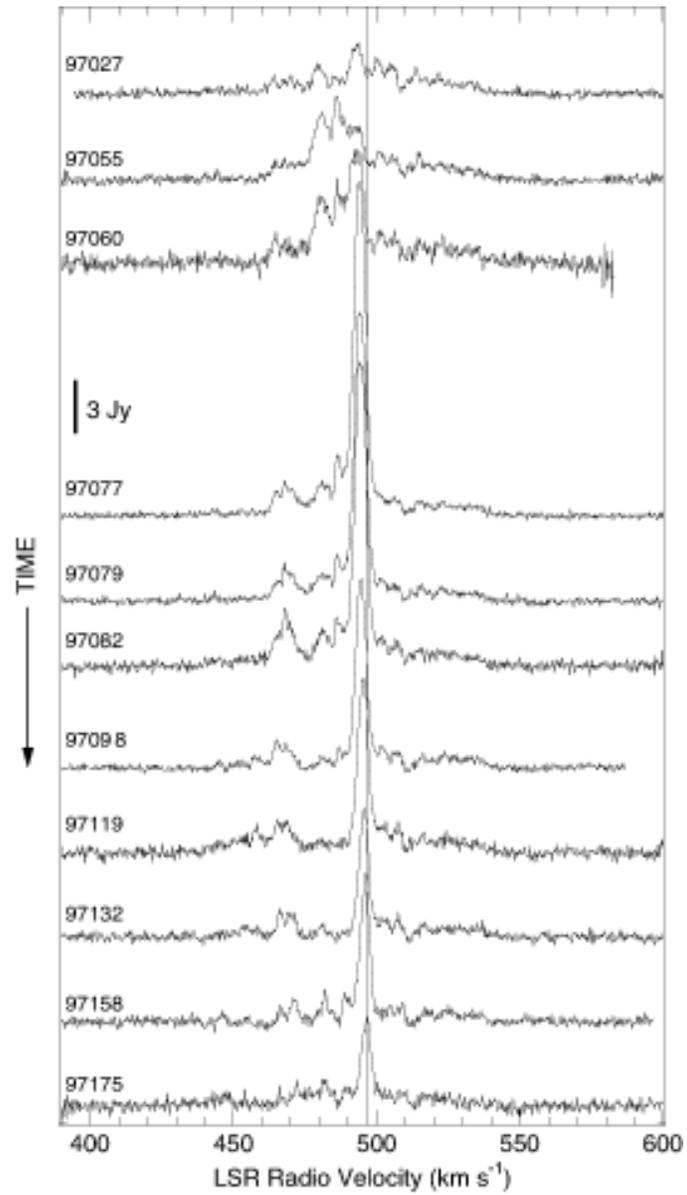}
\caption{Time series of Haystack single-dish spectra of systemic 
maser emission.  Labels indicate the epoch of observation ($yyddd$). A
vertical line marks approximately the velocity of the flaring spectral
feature at the end of the monitoring, $\sim 2$~\kms~above its initial
velocity.}
\label{fg:flarespec}
\end{figure}

\begin{figure}[ht]          
\centering
\includegraphics[width=3.5in]{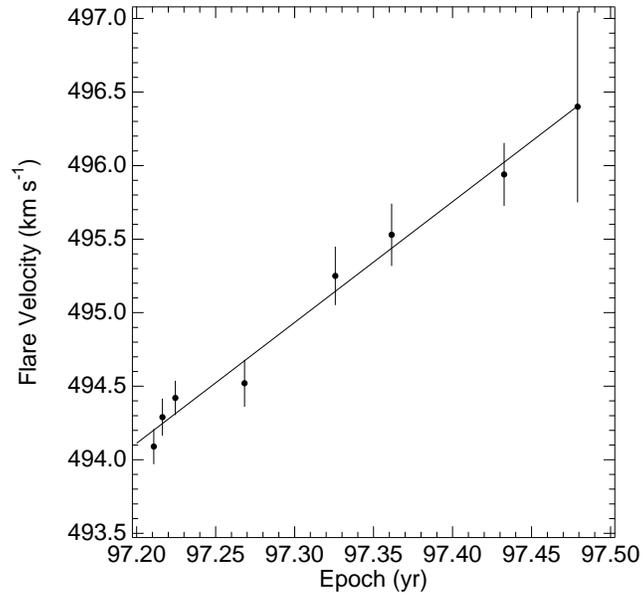}
\caption{Line center velocity of the 1997 flare as determined from
measurements taken at the Haystack Observatory.}
\label{fg:flaredrift}
\end{figure}

\begin{figure}[ht]          
\centering
\includegraphics[width=3.5in]{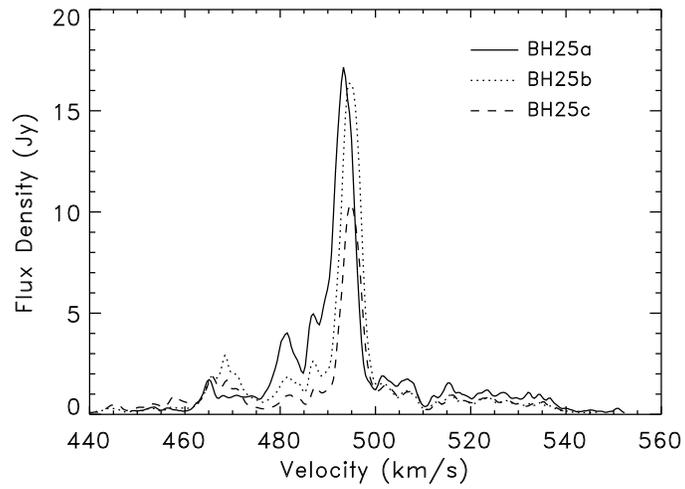}
\caption{Systemic spectra from VLBA epochs BH25a, BH25b, and BH25c
showing the 1997 flare.  These were observed on days 66, 82, and 97 of
1997, respectively.}
\label{fg:bh25_spec}
\end{figure}

The 1996 systemic maser flare arose at a velocity of about
494 \kms\ (Bragg \etal\ 2000).  Figure~\ref{fg:flarespec} shows the
single-dish Haystack detection of a second flare in the NGC 4258
systemic spectrum.  The flare was first unambiguously evident on day
77 of 1997, at which point it was near its maximum flux density of
about 17 Jy.  It returned to a more standard strength of $\sim5$ Jy by
day 175 of 1997.  The implied lifetime of this flare is thus $\la100$
days, somewhat longer than the 75-day upper limit indicated by the
Bragg \etal\ (2000) observations of the 1996 flare.  The centroided
velocity of the flare reddens from 494.1 to 496.4 \kms\ over the
course of the observations, and is well-fit by a secular drift of
$8.21\pm0.82$\,\kmsyr, which is consistent with previous detections of
accelerations in the systemic masers (see Figure~\ref{fg:flaredrift}).
Coincidentally, VLBA epochs BH25a, b, and c (Table~1) occurred near
the maximum of the 1997 flare and detected a peak flux density of 17
Jy and flare velocities between 493.3 and 495.0 \kms\
(Figure~\ref{fg:bh25_spec}).

\begin{figure}[ht]          
\centering
\includegraphics[width=4in]{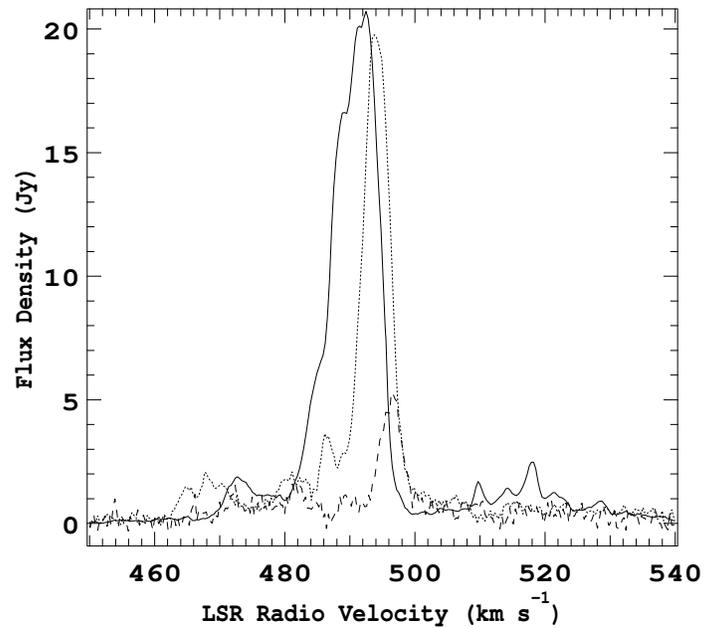}
\caption{Systemic spectra of the 1996 and 1997 flares. The solid line
shows the 1996 flare as detected by the VLA on day 89 of 1996 (from
Bragg \etal\ 2000).  The other two spectra show single-dish detections
of the 1997 flare at the start (dotted) and end (dashed) of its light
curve, on days 77 and 175, respectively, of 1997.}
\label{fg:flare_both}
\end{figure}

Figure~\ref{fg:flare_both} shows spectra for both the 1996 and 1997
flares.  The flare flux density maximums occur between 492.6\,\kms\
(in 1996) and 494.6\,\kms\ (at the end of the 1997 flare).  The line
profile of the 1996 flare was in fact not well-fit by a single
feature.  In a three-feature fit, the two features significantly brighter
than standard systemic emission are centered about 488.4\,\kms\ (14
Jy) and 492.6\,\kms\ (20 Jy).  The impressive agreement in the
velocities of the two flares suggests that the NGC 4258 flare is
actually a recurring event at a particular point in the disk.  That
is, flaring is observed when a suitable patch of enhanced density gas
moves through a zone of optimum amplification. For $\xo=-0.19$\,mas
and $\vo=473.5$\,\kms, Figure~\ref{fg:3line} shows that the flare
position corresponds with the velocity convergence point of the disk
20\,\kms\ and 0.1\,mas to the red of disk center.

Hence, we propose that the NGC 4258 flare is recurring, and arises as
a result of the bowl-like geometry along the front of the disk.  At
$\vlos\approx\vo+20$\,\kms\ the inclination and position angle warps
conspire to bring the LOS velocity gradient along the LOS to
$\la2$\,\kms\ across the entire $\sim2$\,mas width of the bowl. {\it
We speculate that flares arise as a result of the serendipitous
alignment and coherent self-amplification of masers along the bowl
bottom at $\sim\vo+20$\kms.} The time scale between flares of about
one year suggest that the characteristic length scale between gas
density peaks is about $2 \times 10^{16}$ cm.  We note that Boboltz
\etal\ (1998) have shown that a 1990 flare in the W49N water maser is
well-modeled by the same type of chance alignment proposed here.

The inclination and position angle warps indicated by the positions
and velocities of the high-velocity masers unambiguously predict a
velocity convergence point along the near edge of the disk.  However,
the preferred \xo\ and \vo\ of the high-velocity fits do not lead to a
precise coincidence between the velocity convergence point and the
location of the flare. The final entry of Table~2, labeled \didrp, is
the best-fitting inclination-warp model that also leads to center
coordinates consistent with the flare.  It is arrived at using the
following iterative scheme. The high-velocity fitting is first
performed with no constraints on \xo\ or \vo, yielding the solution
\didr.  The velocity convergence point, which depends on the precise
shape of the warp, the central mass, and the disk center coordinates,
is computed for this model.  \xof\ and \vof, the disk center
coordinates that locate the flare at the velocity convergence point
while holding all other model parameters fixed, are estimated, and the
high-velocity fit is rerun around \xof\ and \vof\ priors.  The
sequence is repeated with increasingly narrow prior distributions
about \xof\ and \vof\ until convergence is achieved. In this case,
convergence indicates the existence of an \xof\ and \vof\ for which
the best-fitting central mass and warp model (as determined from the
high-velocity masers) also leads to an alignment between the velocity
convergence point and the flare. 

\begin{figure}[ht]          
\centering
\includegraphics[width=5in]{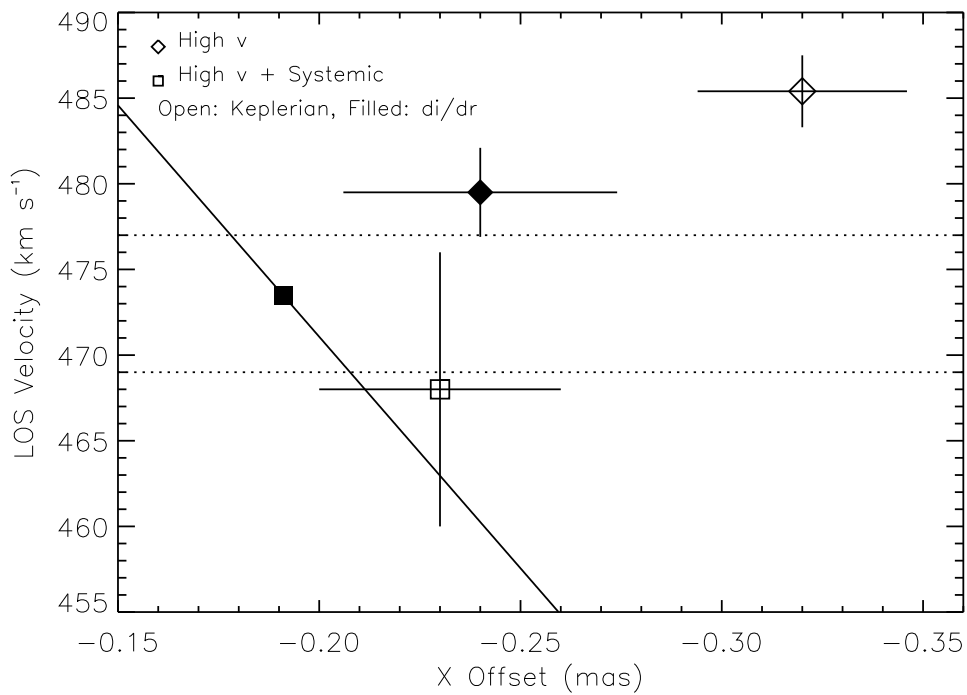}
\caption{Model disk center parameters for a variety of fits.  The
diamonds mark the disk center from fits incorporating only the
high-velocity data: open for a purely Keplerian model, filled for an
inclination-warp model.  The squares are derived from fits
incorporating both the high-velocity and systemic data. The open
square is from the global fit of Herrnstein (1997).  The solid square is
derived from the flare-constrained model of
Section~\ref{sec:flare}. The solid line marks the locus of disk
centers for which the velocity convergence point lies along the
observed systemic emission. The dotted lines demarcate the $1\sigma$ 
confidence interval for the galaxy systemic velocity as measured 
from optical observations.}
\label{fg:center}
\end{figure}

Table~2 shows that there is a $\sim2\sigma$ discrepancy between the
unconstrained high-velocity fit and the flare-constrained fit. There
is also a significant degradation in the reduced \chisq\ of the
flare-constrained fit.  Past efforts at global disk fitting have
uncovered comparable inconsistencies between the systemic and high
velocity masers, and this is summarized in Figure~\ref{fg:center}.
The diamonds denote the maximum-likelihood centers for fits
considering only the high-velocity masers.  The squares mark the
center values for fits incorporating both the systemic and high-velocity
masers. The open and filled symbols represent position-angle warp and
inclination- and position-angle warp models, respectively. The open
square is derived from true global fit incorporating all masers,
while the filled square corresponds to the \didrp\ solution discussed
above, and utilizes the high-velocity masers and the location of the
flare.  The discrepancy between the open diamond and open squares
illustrates the tension between the high-velocity and systemic masers
in the context of a position-angle warp.  The solid symbols indicate
that by including an inclination warp and locating the flare at the
velocity convergence point of the disk, we have significantly reduced,
but not removed this tension.

\subsection{Discussion} 

\begin{figure}[ht]          
\centering
\includegraphics[width=5in]{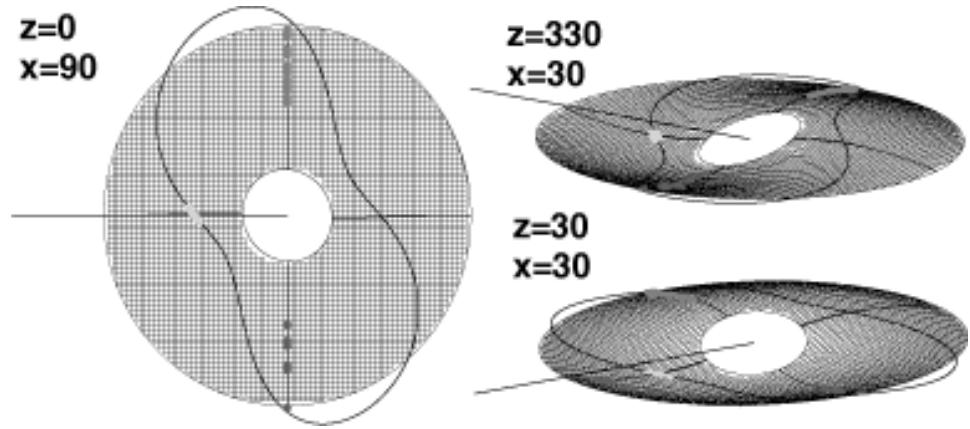}
\caption {Three perspectives of the best-fitting disk incorporating
both a position angle warp and an inclination warp with
$\didr=0.034$\,\rpm.  The line of sight is shown extending beyond the
outer edge of the disk.  The disk midline, the $\theta=0^{\circ}$ and
$\theta=270^{\circ}$ radials, and the locus of disk tangents are
overlaid on the disk.  The masers, from all the epochs in Table~1,
have been positioned according to their observed $x$ and $y$, and
assuming that the high-velocity features lie along the midline and the
systemic features lie along the bowl bottom.}
\label{fg:mesh}
\end{figure}

In this section we have attempted to demonstrate that the inclination
warp derived from the projected high-velocity rotation curve naturally
explains a number of otherwise puzzling features of the systemic
masers.  In particular, we have shown that the bowl geometry provides
an explanation for the extremely tight declination spread, and broader
velocity spread of the systemic masers. Furthermore, we have argued
that the combination of central X-ray heating, background
amplification, and thermal absorption makes the bowl bottom an
especially attractive location for the systemic masers to arise.
Finally, we have speculated that the apparently recurring flare in
NGC~4258 arises at the point in the bowl where the velocity gradient
along the LOS approaches zero.  {\it Based on these results, we
conclude that $\didr=\didrp=0.034$\,\rpm\ and that the inclination
warp is in fact predominantly responsible for the flattened
high-velocity projected rotation curve.}  Figure~\ref{fg:mesh} shows
three perspectives of this disk. The disk midline and the
$\theta=0^{\circ}$ and $\theta=270^{\circ}$ radials are overlaid.
The peanut-shaped curve traces the location of the disk tangent, as
seen from our perspective.  As prescribed by the model, the systemic
masers have been distributed along the disk tangent at the near edge
of the disk, while the high velocity masers have been placed along the
midline.

\section{Implications}

\subsection{The Rotation Curve Revisited}

Of the three specific refinements to a constant-inclination, centrally
bound disk considered in Section~\ref{sec:refine}, it was the
inclination warp alone that was capable of steepening the rotation
curve and that prevented us from placing interesting limits on $\xi$,
\rhoc, and \mdota.  The dotted curves in Figure~\ref{fg:chisq} show
$\chisq(\xi)$, $\chisq(\rhoc)$, and $\chisq(\mdota)$ for a disk with
$\didr=0.034$\,\rpm. Figure~\ref{fg:pdfs_pert} also shows the
probability distribution for $\xi$ in the presence of an inclination
warp. {\it The maximum-likelihood rotation law for a
inclination-warped disk is $\vrot\propto r^{-0.49\pm0.01}$, and there
is no significant evidence for deviation from pure-Keplerian
rotation.}  The asymptotic behavior of the relative \rhoc\ and \mdota\
\chisq\ further reinforces the conclusion that the inclination warp
alone fits the projected high-velocity rotation curve. For
$\didr=0.034$\,\rpm, we find that $\rhoc\ga10^{12}$\,\smpc\ and
$\mdot\la5\times10^{-2}\alpha$\,\smy.

Maoz (1998) has developed a formalism for estimating cluster lifetime
upper limits (\taumax) by considering both collision and evaporation
timescales for a variety of cluster constituents.  For NGC 4258, he
finds $\taumax\la1$\,Gyr for $\rhoc\ga8\times10^{11}$\,\smpc.  Hence,
the lower limit on \rhoc\ derived from the high-velocity rotation
curve indicates that if the disk is bound by a central cluster, the
cluster lifetime would have to be considerably shorter than the age of
the galaxy itself. This strongly favors the presence of a supermassive
black hole in NGC 4258. The upper limit on \mdota\ is less interesting
since it is three times larger than what would be expected for
advection-dominated accretion flow models (ADAF; Gammie, Narayan, \&
Blanford 1999), which are generally the least efficient, highest
\mdot\ solutions.

\subsection{The Disk Geometry Revisited}

For a disk with no position angle warp, the narrow spread in the
declinations of the systemic masers implies that these features are
tightly clustered about a radius of 0.14\,pc (3.9\,mas).  A number of
theories have been proposed to explain the inner cutoff in the
systemic emission.  Neufeld \& Maloney (1995) suggest that the inner
edge of systemic emission might be due to a flattening of the disk and
a resulting decrease in the X-ray heating inside of this radius.  More
recently Yuan \etal\ (2002) have fit the NGC 4258 SED with a
jet-dominated AGN model in which the outer thin disk transitions to an
ADAF at a radius \rtrans.  The NGC 4258 SED alone does not place
tight constraints on \rtrans, and it is tempting to associate \rtrans\
with the inner edge of systemic emission.  However, if the inclination
warp model is correct then the systemic masers are clustered along the
bottom of a bowl along the near edge of the disk, and there is no
additional explanation needed for their narrow spread in $y$.  In this
model, there is nothing distinctive about 0.14\,pc except that {\it
from our perspective} it corresponds to a tangent point to the disk.

\begin{figure}[ht]          
\centering
\includegraphics[width=5in]{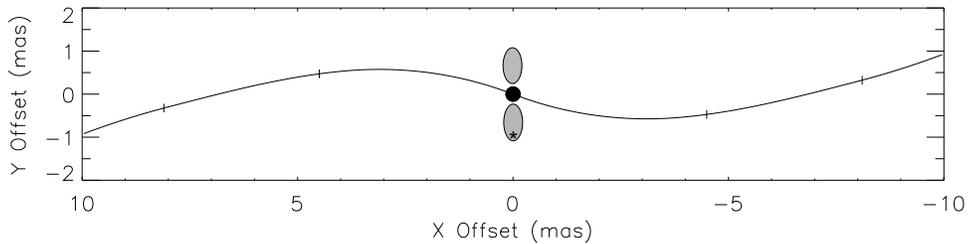}
\caption{Cross-section of the disk in the plane of the sky, along the
disk midline, showing the position angle warp. The warp has been fit
using VLBA epochs BM19, BM36a, BM36b, and BM56b, and is given by
$\phi=0.425-0.088r + 0.0038r^2$.  The small vertical dashes mark the
envelope of observed high-velocity maser emission. The northern
shaded oval marks the observed location of northern jet emission, and
the southern oval is the reflection of this through the disk center.
The asterisk marks the observed location of the southern jet
emission.}
\label{fg:pslice}
\end{figure}

Similarly, efforts have been made to explain the observed range of
high-velocity emission from 0.16 to 0.28~pc (4.5--8.1\,mas) on either
side of the central mass.  Most notably, Neufeld \& Maloney (1995)
argue that the outer edge of high-velocity emission is due to a
transition from a molecular to an atomic disk at this radius.  We
discuss this in detail in the next section.  To date there is no
compelling explanation for the inner edge of high-velocity emission.
Here we speculate that the same amplification and absorption
considerations that make the bowl model attractive as an explanation
for the systemic geometry play a role in the inner cutoff for the
high-velocity emission.  Figure~\ref{fg:pslice} shows the disk cross
section in the plane of the sky.  The vertical dashes mark the
observed extent of high-velocity emission.  Water molecules inside of
a certain radius will have an unabsorbed view of the background jet
emission and, according to the arguments of Section~\ref{sec:bowl},
will generate strong maser emission perpendicular to our LOS (i.e.,
``their'' systemic emission). If these masers are at least mildly
saturated, as appears to be the case for 'our' systemic masers
(e.g. Modjaz et al. 2005), then competitive gain arguments would
suggest a corresponding quenching of any high-velocity emission along
our LOS from within this radius.  The precise radius at which the
background jets come into view depends on the orientation and extent
of the jets. But from Figure~\ref{fg:pslice}, which shows the observed
extent of continuum emission, it seems plausible that there is indeed
a connection between the purported location of 'their' systemic
emission and 'our' high-velocity emission.

\subsection{X-ray Obscuration and the Accretion Rate \label{sec:xray}}

For $\didr=0.034$\,\rpm, the accretion disk traced by the masers rises
in front of and obscures the central engine at a disk radius, \rx, of
8.3\,mas or 0.29~pc (see Figure~\ref{fg:cross_bowl}), where
$\rx=i'_o/(\didr)$ and $i'_o$ is the inclination at $r=0$.  The formal
fractional uncertainty in this radius is about 30\% and is dominated
by the uncertainty in \didr. However, for a disk in which the systemic
masers lie along a tangent to the disk, \rx\ can be written in terms
of \rs\ and \is\ instead of $i'_o$ and \didr\ as
$\rx=\rs[1+\is/tan(\is)]$, where \rs\ and \is\ are the radius and
inclination ($0^{\circ}$ for edge-on) of the systemic masers.  In
other words, if we {\it assume} that the systemic masers are clustered
around the bottom of a bowl along the near edge of the disk, then the
X-ray obscuration radius is independent of \didr\ and depends only on
other relatively well-constrained parameters of the model.  We
estimate that in the context of the bowl model, \rx\ is known to a
fractional uncertainty of about 5\%.

From Equation \ref{eq:centraldensity}, the shielding column density
through a Shakura-Sunyaev isothermal accretion disk is
\bq
\ns=1.3\times10^{-3}\frac{\mdotmf\masse^{1/2}}{\alpha\mu\rpc^{3/2}\css^{2}}~~{\rm
cm}^{-2}.
\label{eq:col}
\eq Here, \ns\ is the shielding column in units of
$10^{24}$\,cm$^{-2}$, $\mdotmf=10^{5}\mdot$, $\masse=10^{-8}\mo$,
\rpc\ is the disk radius in pc, and $\css=\cs/7$\,\kms\ is the
isothermal sound speed. The column is computed for an obliquity angle
$\cos^{-1}\mu$ with respect to the disk normal
($\mu=\nhat(r,\theta)\cdot\rhat$) and assuming solar abundances.  From
the bowl model the obliquity angle is 74 degrees and $\mu=0.27$.
Neufeld \& Maloney (1995) showed that as a result of the warp in the
NGC~4258 disk, heating from direct exposure to central X-rays
dominates viscous heating in the outer disk, and this must be taken
into consideration in evaluating Equation~\ref{eq:col}.  More
precisely, they showed that for an externally heated Shakura-Sunyaev
disk, the midplane pressure scales as $r^{-3}$. As a result there
exists a critical radius, \rcr, beyond which the disk is atomic
throughout: \bq
\rcr=0.04\lum^{-0.426}(\mdotmf/\alpha)^{0.809}\mu^{-0.383}\masse^{0.617}\css^{-1.19}\mbox{
pc},
\label{eq:rcr}
\eq
where \lum\ is the 2--10 keV X-ray luminosity in units of
$10^{41}$\,\ergs.  Numerical simulations by Neufeld, Maloney, \&
Conger (1994) indicate that the temperature in the predominantly
atomic disk beyond \rcr\ rapidly reaches an asymptotic value of about
8000\,K, corresponding to $\css=1$.  Their models also indicate that
within \rcr, the temperature of the dwindling surface atomic layer
ranges from about 5000 to 8000\,K degrees depending on depth and
proximity to \rcr, while the midlplane molecular layer temperature
ranges from about 400--1000\,K.

Equations~\ref{eq:col} and \ref{eq:rcr} can be used to estimate the
X-ray shielding column at \rx\ as a function of $\mdotmf/\alpha$. For
each accretion rate, Equation~\ref{eq:rcr} is first used to calculate
\rcr\ along the $\theta=270^{\circ}$ radial for $\css=1$, $\lum=0.8$
(Fruscione \etal\ 2005), and $\didr=0.034$\,\rpm. If for a given
$\mdotmf/\alpha$, $\rx>\rcr$, then the disk is entirely atomic at \rx\
and \ns\ is computed from Equation~\ref{eq:col} using the asymptotic
sound speed of 7\,\kms\ and $\mu=\mu_{rx}(\approx 0.27)$.  If on the
other hand $\rx<\rcr$, then the X-rays must also transit cooler
molecular gas in the disk midplane.  

\begin{figure}[htb]          
\centering
\includegraphics[width=5in, angle=90]{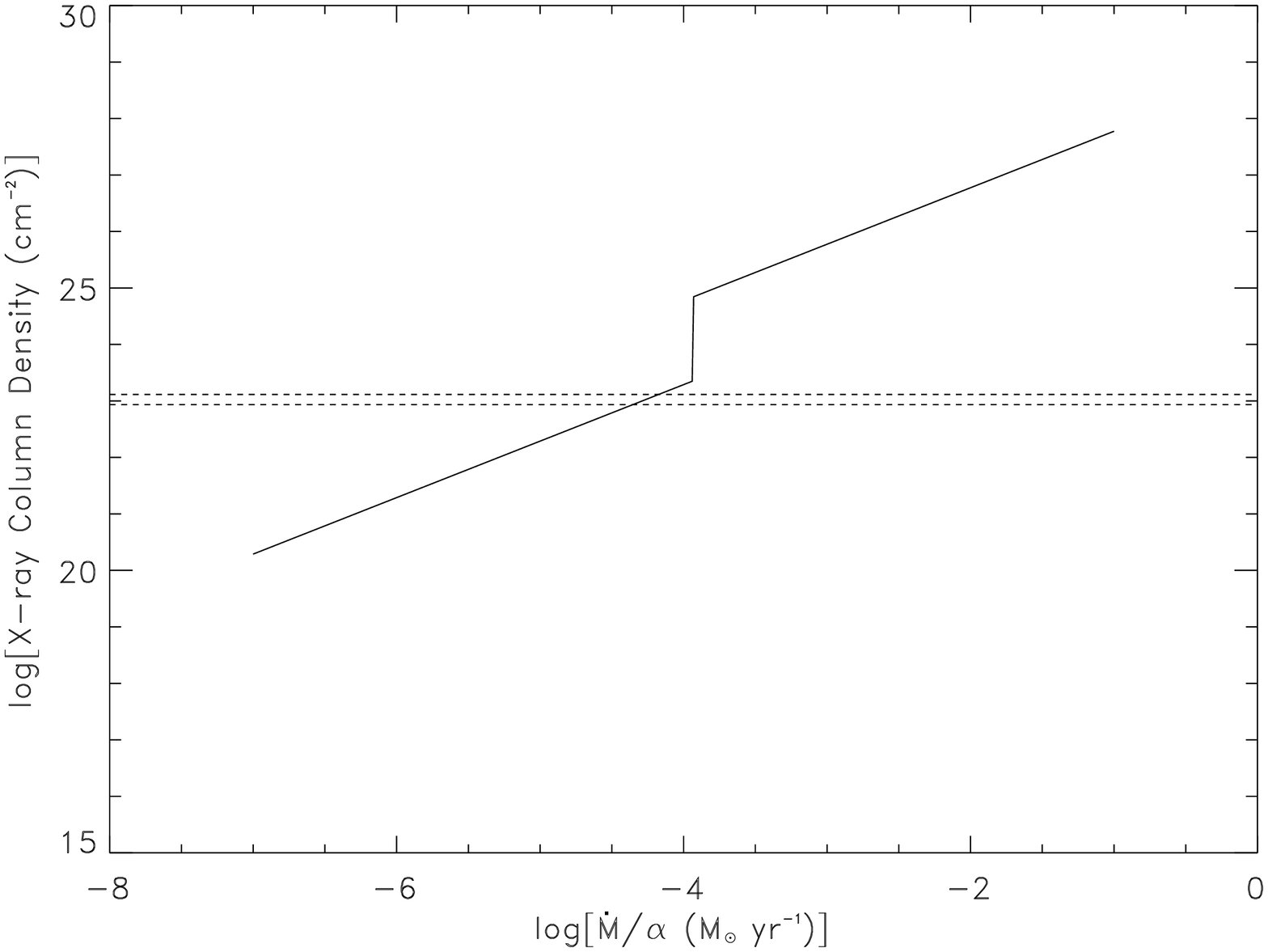}
\caption {X-ray column density through the disk as a function of
\mdota\ for $\lum=0.8$.  The column is calculated at a radius, \rx, of
8.3\,mas (0.29~pc), at which point the disk rises in front of the
central X-ray source for $\didr=0.034$\,\rpm. For $\mdot\la
10^{-4}\alpha$\,\smy, the disk at \rx\ is atomic throughout.  As the
accretion rate passes through $10^{-4}\alpha$\,\smy\ the disk at \rx\
transitions to cool, molecular gas, and this leads to the sudden jump
in the column.  The two dotted lines mark the range of observed X-ray
columns in the {\it XMM} data of Fruscione \etal\ (2005), and provide
an upper limit to the disk column.}
\label{fg:xrays}
\end{figure}

Figure~\ref{fg:xrays} shows $\ns(\mdota)$ at $\rx=8.3$\,mas.  The
abrupt jump in \ns\ at $\mdot\approx10^{-4}\alpha$\,\smy\ marks the
accretion rate at which \rcr\ passes outside of \rx\ and the disk
column becomes increasingly dominated by cool, dense molecular gas in
the disk midplane.  We have made no effort to model the transition
from predominantly molecular to predominantly atomic gas. However, we
note in passing that the actual transition is likely to be extremely
rapid as a result of the strong radial dependence ($r^{-3}$) of the
disk midplane pressure.  Furthermore, Figure~\ref{fg:xrays} is based
on the assumption that inside of \rcr, all the disk mass is molecular.
While a surface layer of atomic gas will in fact persist inside the
critical radius, it is reasonable to approximate that the column will
rapidly become doiminated by the dense midplane molecular gas.

The X-ray column has been observed and estimated numerous times since
1993 (see Table~2 of Fruscione \etal\ 2005 for a summary), and the
dotted lines in the figure show the range of observed values.
Standard AGN models suggest that this column is due to a thick dusty
torus 1 to 10 pc from the central engine.  With this in mind, we treat
the observed column as an upper limit on the column through the
sub-parsec disk.  For $\rcr>\rx$ the disk column is $50-100$ times
greater than the largest observed column.  {\it Hence, for
$\didr=0.034$\,\rpm, consistency with the observed X-ray column
requires that the disk be atomic at the obscuring radius of 8.3\,mas\
(0.29\, pc), translating to an upper limit on the accretion rate of
$\sim10^{-4}\alpha$\,\smy\ for $\css=1$ and $\lum=0.8$.}  As discussed
above, these conclusions are robust to uncertainties in $\didr$ if the
systemic masers are indeed distributed about the tangent to the disk.
The upper limit on \mdot\ assumes a disk filling factor of order unity
at \rx.  The unbroken nature of the systemic spectrum suggests this is
a reasonable approximation at $\sim4$\,mas.  Furthermore, the relative
stability (to factors of a few; Fruscione \etal\ (2005)) of the
observed column over the last ten years implies a fairly smooth
intervening medium.  Finally, our analysis implicitly treats the
central X-ray emission as arising from a point source.  Observed X-ray
variability on timescales of hours (Fiore \etal\ 2001) implies an
X-ray source of less than 100\,\rg\, corresponding to an angular size
of $\sim10'$ at the distance of the innermost maser.  Thus, these
conclusions should not be sensitive to spatial structure in the
central X-ray source.

\begin{figure}[htb]          
\plotone{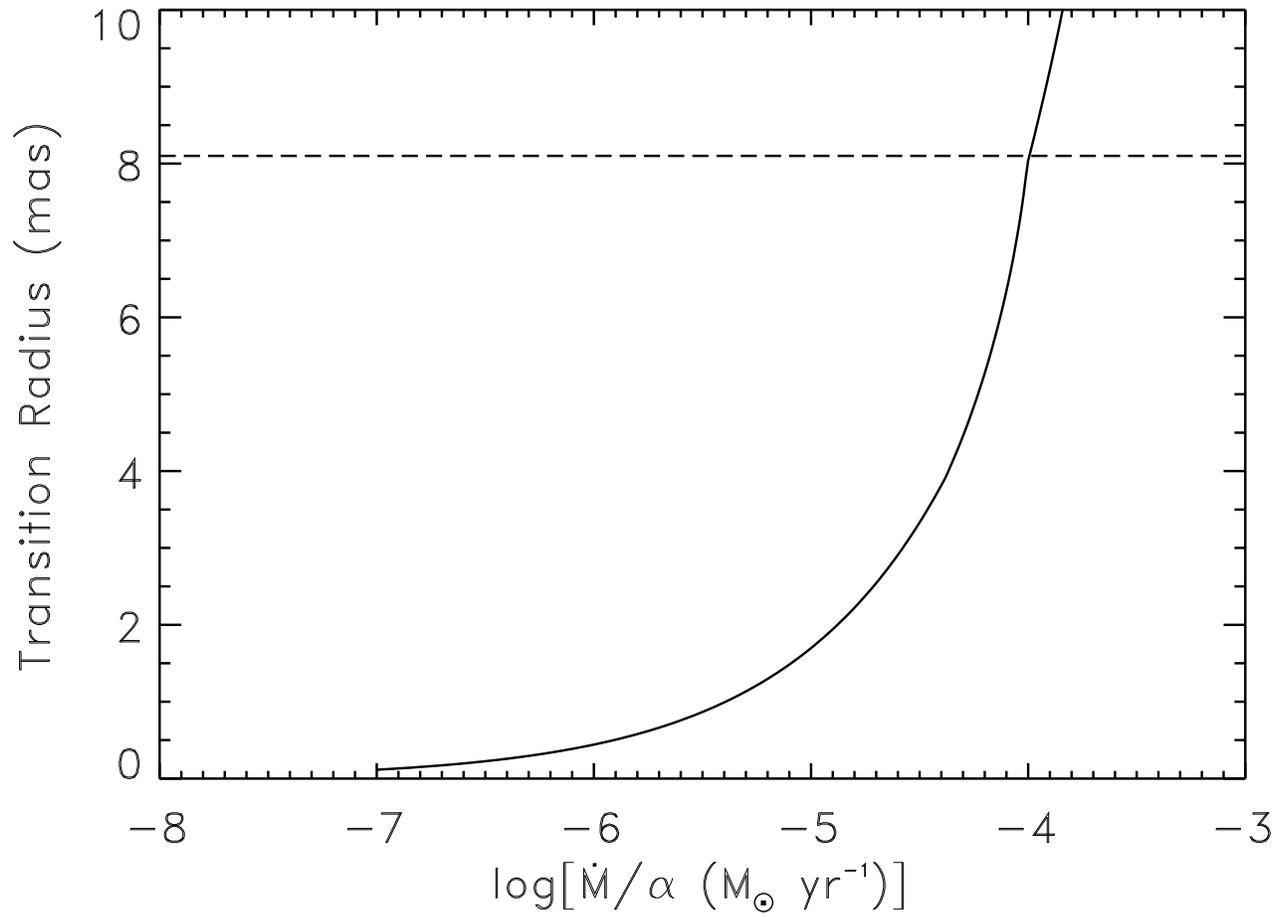}
\caption {Atomic-to-molecular phase transition radius along the disk
midline as a function of \mdota\ for $\css=1$ and $\lum=0.8$.  The
dashed line corresponds to the radius of the outermost high-velocity
maser, and provides a lower limit to the actual phase transition
radius.}
\label{fg:rcrit}
\end{figure}

Neufeld \& Maloney (1995) estimate the accretion rate through the
NGC~4258 outer disk by postulating that it is the transition from a
molecular to an atomic disk at \rcr\ that defines the radius, \router,
of the outermost high-velocity maser.  Using Equation~\ref{eq:rcr} and
setting $\rcr=8.1$\,mas they find $\mdot=7\times10^{-5}\alpha$\,\smy.
A more conservative approach is to require merely that the disk be
molecular over the entire range of maser emission, leading to a lower
limit on the accretion rate.  Figure~\ref{fg:rcrit} shows
$\rcr(\mdota)$ along the midline for the position angle warp discussed
in the previous section. {\it In order to accommodate both the
manifest existence of maser emission at \router\ and the observed
X-ray column at \rx, the transition from molecular to atomic gas must
occur outside of 8.1\,mas (0.28\,pc) along the disk midline and inside
of 8.3\,mas (0.29\,pc) along the LOS.  For $\css=1$ and $\lum=0.8$,
this implies accretion rate upper and lower limits of
$\sim10^{-4}\alpha$\,\smy.} Given the central mass of
$3.79\times10^{7}$\,\msun, this translates to an accretion rate in
Eddington units of $\sim10^{-4}\alpha$, assuming a canonical
efficiency of 0.1.  For $\lbol\sim10^{42}$\,\ergs, we estimate a
radiative efficiency of $\ga0.2$ for $\alpha\la1$.

The conclusion that the disk transitions from molecular to atomic gas
between \router\ and \rx\ is robust in the sense that a predominantly
molecular disk would lead to substantially more column than observed
and an atomic disk could not support masers.  Furthermore, our
estimate that the transition occurs at $\sim8.2$\,mas$=0.28$\,pc is
relatively insensitive to details of the fitting process as long as it
is indeed the case that the systemic masers lie along the bottom of a
bowl and the high-velocity masers trace a position angle warp.  The
accretion rate estimate, on the other hand, is sensitive to estimates
for the X-ray luminosity and column (both known to be variable by
factors of a few; Fruscione \etal\ 2005), the disk sound speed, and,
more generally, to the applicability of the equations for an
externally heated thin disk to the outer NGC 4258 disk.  With this in
mind, we refrain from placing formal errors on the accretion rate.

The preceding analysis validates the proposal of Neufeld \& Maloney
(1995) that the high-velocity emission is truncated at large radii by
a transition from molecular to atomic gas.  In addition, it suggests
that a substantial fraction, if not all, of the observed X-ray column
is due to the warped disk at 0.29\,pc (Figure~\ref{fg:xrays}).  The
standard model of low-luminosity AGN (Antonucci \& Miller 1985)
asserts that Seyfert 1s and 2s are the same class of object observed
from different perspectives.  More recent work suggests that HBLR and
non-HBLR Seyfert 2s may in fact be distinct classes of objects, with
the HBLR Seyfert 2s being the misaligned counterparts to the face-on
Seyfert 1s (c.f. Veilleux 2003).  While it is generally accepted that
at least some Seyfert 2s are Seyfert 1s shrouded in intervening gas,
the precise nature of the absorbing material is still debated.  Broad
emission-line region clouds, sub-parsec-scale warped disks, 1--10\,pc
thick molecular tori, and larger-scale wind structures have all been
considered.  Recent studies of column variability in Seyfert 2
galaxies suggest intervening material quite close to the central
engine (Risaliti, Elvis, \& Nicastro 2002).  NGC~4258 is just one
object, and as such generalizations cannot be drawn from it. {\it
  Nevertheless, our analysis, together with the X-ray observations of
  Fruscione \etal\ (2005), presents compelling evidence for at least
  one HBLR Seyfert 2 in which most, if not all, of the X-ray
  obscuration arises in a subparsec-scale warped disk.}

If the X-ray column indeed arises in the disk, then the expected
timescale, \tvar, for variability in the X-ray column is $H/\vrot$,
where we have associated the disk thickness with the characteristic
scale of the intervening material. For an isothermal disk,
$H=r\cs/\vrot$, and $\tvar=r\cs/\vrot^{2}$.  For the NGC~4258 disk at
0.29\,pc, $\vrot\simeq750$\,\kms\ and, taking $\cs=7$\,\kms, we
estimate $\tvar\approx3.6$\,yrs.  The observed column variability of
$\sim60\%$ over $\sim5$ months (Fruscione \etal\ 2005) is
approximately consistent with this prediction, but may indicate
inhomogeneities on a scale somewhat less than the disk thickness.

The existence of water masers requires $300\la\tm\la1000$\,K and
$10^{8}\la\rhom\la10^{10}$\,\percmcmcm\ in the NGC 4258 disk, where
\tm\ and \rhom\ are the temperature and density of the masing
gas. This temperature is consistent with the models of Neufeld \etal\
(1994) for an externally irradiated thin disk.  The density of the
maser layer is estimated by {\it assuming} that the masers lie
approximately at the midplane of the disk.  From
Equation~\ref{eq:centraldensity} the midplane number density, $n_{H2}
= \rho_{mid}/m_{H2}$ can be written 
\bq n_{H2} = 3700
\frac{\mo\mdot}{r^3\sound^3\alpha} \mbox{ \percmcmcm}.
\label{eq:lastone}
\eq For $\mdot=1\times10^{-4}\alpha$ \smy\, and $\sound = 2.3$ \kms
($\tm=800$ K), $n_{H2}$ ranges from $7\times10^{8}$\,\percmcmcm\ at
the inner radius of 0.14~pc to $0.9\times10^{8}$\,\percmcmcm\ at
0.28~pc, the radius of the outermost high-velocity feature.  Hence,
the accretion rate derived from the disk geometry and X-ray column
provides conditions suitable for water masers in the disk midplane
over the entire envelope of observed maser emission.

The same disk material that obscures the central X-ray emission will
also obscure any radio emission arising in the central
engine. Herrnstein \etal\ (1998) report a 220\,$\mu$Jy upper limit on
22~GHz emission coincident with the central mass, and this limit has
been used to place constraints on the accretion mode and rate through
the disk (Herrnstein \etal\ 1998; Gammie, Narayan, \& Blandford 1999;
Yuan \etal\ 2002).  The 22-GHz optical depth due to free-free
absorption in the disk at \rx\ will be 1--2 (Herrnstein, Greenhill, \&
Moran 1996), and hence we speculate that the actual upper limit on
central-engine 22~GHz is likely 1--2 times larger than the observed
220\,$\mu$Jy.

Finally, we note that any association between the accretion rate in
the maser disk and that in the NGC 4258 central engine implicitly
assumes that mass loss via winds can be ingored and that the disk is
in a steady state, such that $\partial\mdot(r)/\partial r=0$.  Gammie
\etal\ (1999) argue that in the absence of external torques, steady
state is achieved after a viscous timescale given approximately by
$\tv\simeq0.26\alpha^{-1}(200\mbox{ K}/T)(r/0.2\mbox{
pc})^{1/2}$\,Gyr.  For $T=\tm=800$\,K and $\alpha\ga0.3$,
$\tv\la0.2$\,Gyr at the radii of the masers.

\subsection{The NGC 4258 Central Engine \label{sec:engine}}

Ho (2003) lists the defining properties of Low luminosity AGN.  LLAGN
are highly sub-Eddington, and their spectral energy distributions
(SEDs) lack the 'big blue bump' of more powerful AGN.  Instead they
typically show an excess of IR emission, perhaps even peaking in this
region of the spectrum. They tend to be relatively strong radio
emitters, and they possess narrow rather than relativistically
broadened iron k$\alpha$ emission.  NGC 4258 satisfies all of these
criteria, especially if we accept that the steep ($s=1.1\pm0.2$)
optical/UV spectrum of Wilkes \etal\ (1995) is in fact an extension of
the steep ($s=1.4\pm0.1$) IR spectrum of Chary \etal\ (2000), and that
the SED therefore peaks in the IR.

The generic LLAGN SED as described above shows indications that at
least three distinct emission mechanisms may be at work.  First, the
IR excess (as well as numerous examples of double-peaked emission
lines; Ho \etal\ 2000) is consistent with black body emission from a
standard optically thick, geometrically thin, radiatively efficient
Shakura-Sunyaev disk.  However, the absence of an optical/UV peak and
of relativistically broadened k$\alpha$ lines indicates this disk does
not extend too close to the black hole.  Next, the representative LLAGN
SED seems broadly consistent with emission from a class of hot,
quasi-spherical, inefficient accretion flows, known collectively as
radiatively inefficient accretion flows (RIAFs; see Quaetert 2003 for
a recent summary).  Accretion of this type naturally explains the low
bolometric luminosities and inferred low accretion efficiencies of
LLAGN. Finally, the characteristic radio excess is most readily
explained by compact jet emission at the central engine (Falcke \&
Biermann 1995, 1999; Anderson \etal\ 2004).

Hence, a plausible model for the LLAGN central engine (c.f. Ho 2003;
Qataert 2003) has an outer thin accretion disk truncated at
$\sim100-1000$\,\rg\ by a hot, inefficient RIAF, which in turn
channels some fraction of the accreted energy into a compact jet.
While it is perhaps uncontrovertial that one or all of these
mechanisms play a role in most LLAGN central engines, the neat
itemization of the previous paragraph is probably misleading.  The
RIAF models can be enhanced to accommodate the radio excess, the
compact jet models can be extended to accommodate essentially the
entire SED, and the thin disk can be left out altogether.

Several attempts have been made to model the NGC~4258 spectrum.
Gammie \etal\ (1999) propose an accretion rate of $\sim0.01$\,\smy\
through an advection dominated accretion flow (ADAF) that transitions
to a thin disk at $10-100$\,\rg. The model relies on relatively high
accretion rates through an outer thin disk to explain the observed IR
luminosity and the extreme inefficiency of the central ADAF to
accommodate the significant sub-Eddington luminosity of the system.
However, this model appears to be inconsistent with the subsequent
detection of a steep ($s=1.4\pm0.1$) IR spectrum by Chary \etal\
(2000).  Yuan \etal\ (2002), on the other hand, propose that
essentially the entire NGC~4258 LLAGN spectrum arises in a compact
jet.  They demonstrate broad agreement with the observed SED, and in
particluar the steep IR spectrum, for jet mass loss rates between
$2.9\times10^{-4}$ and $7\times10^{-5}$\,\smy.  They speculate that
disk accretion rates of $10^{-3}-10^{-4}$\,\smy\ are needed to feed
the jets, and show that for such low accretion rates, an outer thin
disk and central ADAF contribute negligably to the jet sprectrum.  In
this jet-dominated model, the radiatively inefficient ADAF serves
primarily as a means to channel large amounts of accretion energy to
the radiatively efficient jets.  {\it Our estimate for the NGC~4258
accretion rate is in reasonable agreement with the jet-dominated
models of Yuan \etal\ (2002)}.  In NGC~4258 we know that all the
detected 22~GHz emission arises in compact jets and apparently not in
a central RIAF.  Furthermore, we appear to be able to place relatively
tight constraints on the accretion rate at 0.28\,pc, and this may aid
in untangling the relative contributions of jet and RIAF.

\section{Conclusions}

In this paper, we have proposed a geometry for the sub-parsec disk in 
NGC~4258 based primarily on its ability to explain a large number of 
oherwise puzzling features of the system.  In particular, an inclination
warp explains:
\begin{itemize}
\item The sub-Keplerian projected rotation curve of the high-velocity 
masers and implies a purely Keplerian underlying disk.
\item The narrow declination spread of the systemic masers, and their
apparent tight clustering around a radius of $\sim4$\,mas,
corresponding to the bottom of a bowl along the near edge of the disk.
\item An apparently recurring flare in the systemic masers $\sim20$\,\kms\ 
red of the systemic velocity of the galaxy.
\item In conjunction with observed X-ray column, the outer cutoff of 
high-velocity emission.
\item By analogy with amplification and obscuration arguments for the 
systemic masers, the inner cutoff of the high-velocity emission.
\end{itemize}

While the model is compelling in its ability to explain a variety of
aspects of the system, we have not demonstrated that it is the best
model in the formal maximum likelihood sense.  In particular, we have
shown that there remains a $2\sigma$ tension between the disk center
predicted purely from the high-velocity masers, and the center derived
from the high-velocity masers and the flare location.  The resolution
of this tension requires global fits, perhaps incorporating additional
details not considered here (such as non-circular orbits).

Finally, the inclination-warp model makes several predictions, most
obviously that systemic maser emission cannot arise significantly
further south of disk center than 0.55\,mas, and that all
high-velocity emission must occur within $\sim8.1$\,mas. This has held
true up to this point, and will be easily verifiable in upcoming
datasets.  In addition, it predicts that future flares should occur
$\sim20$\,\kms\ to the red of the systemic velocity.  If it is indeed
the case that the inner edge of high-velocity emission is due to
competitive gain with stronger, amplified emission in the plane of the
sky, then this is not a hard edge like the outer cutoff.  To the
extent that the systemic spectrum is variable and occassionally weak
at certain velocities, we might expect to see faint high-velocity
emission inside of the ususal $\sim0.16$\,pc inner cutoff.  Finally,
we point out that in this model there is no {\it a priori} reason that
the systemic masers must lie along the bowl bottom.  Rather, this is
where such emission will be brightest.  We speculate that there could
be additional much weaker low-velocity emission displaced from the
bowl bottom.  The most likely place to find such emission would be
$\sim20$\,\kms\ to the red of the systemic velocity, where the
velocity gradient along the LOS is smallest.

\acknowledgments{We thank R. Herrnstein for constructive discussions
and for her tireless help in bringing this project to completion.}

\begin{table}[htb]   
\begin{center}
\caption{NGC 4258 VLBI Observing Log} \vspace{0.4cm}
\label{tb:tab1}
\begin{tabular}{cccccc} \hline\hline
 Experiment    &  Observation    &  Antennas$^{(1)}$& Frequency &    Synthesized     &  Sensitivity$^{(4)}$    \\ 
 Code               & Date            &                &Setup$^{(2)}$&     Beam$^{(3)}$   &         \\ \hline
   BM19             &   19940426      & V+Y        &     L       & 1.1x0.6 @ $9^{\circ}$   &     6.0         \\
   BM36a            &   19950108      & V+Y        &     L       & 0.8x0.7 @ $-15^{\circ}$ &     6.5         \\
   BM36b            &   19950529      & V+Y        &     L       & 1.0x0.7 @ $-10^{\circ}$ &     5.2         \\
   BM56b            &   19960921      & V+Y+G+E  &     L       & 1.3x1.1 @ $20^{\circ}$   &    3.0         \\
   BH25a            &   19970307      & V+Y+G+E  &     C       & 0.7x0.3 @ $15^{\circ}$  &     3.0         \\
   BH25b            &   19970323      & V+Y+G+E  &     C       & 0.6x0.3 @ $15^{\circ}$  &     2.3         \\
   BH25c            &   19970407      & V+Y+G+E  &     C       & 0.6x0.3 @ $15^{\circ}$  &     2.6         \\
\hline
\end{tabular}
\end{center}
\tablenotetext{1}{V: VLBA; Y: phased VLA; G: NRAO 140-foot; E: Effelsberg 100-meter.}
\tablenotetext{2}{L: Line mode; C: Continuum mode (see text).}
\tablenotetext{3}{mas x mas @ position angle.}
\tablenotetext{4}{mJy/km/s/beam.}
\end{table}

\begin{table}[htb]   
\begin{center}
\caption{Rotation Curve Models} \vspace{0.4cm}
\label{table:model}
\begin{tabular}{l c c c c c c c} \hline\hline
Model & Relativity? & $x_0$\tablenotemark{a} & $v_0$ & $M_0$ & $\chi_r^2$ & Parameter Value & $\Delta v$\tablenotemark{b} \\
 &  & (mas) & (km s$^{-1}$) & ($\times10^7$~M$_\odot$) &  &  & (km~s$^{-1}$)\\
\hline
Keplerian  & No  & $-0.29\pm0.03$ & $488\pm2$ & $3.85\pm0.01$ & 1.24 & ... & ...\\
Keplerian  & Yes & $-0.33\pm0.02$ & $486\pm2$ & $3.84\pm0.01$ & 1.24 & ... & ...\\
Power Law & Yes & $-0.19\pm0.04$ & $475\pm3$ & $3.56\pm0.01$ & 1.09 & $0.48\pm0.01$~\tablenotemark{c} & 9.0\\
Central Cluster & Yes & $-0.19\pm0.05$ & $475\pm4$ & $3.74\pm0.02$ & 1.07 & $5.7(\pm1.4)\times10^{11}$~\tablenotemark{d} & 10.9\\
Massive Disk  & Yes & $-0.20\pm0.04$ & $476\pm3$ & $3.80\pm0.01$ & 1.10 & $6.5(\pm1.3)\times10^{-2}$~\tablenotemark{e} & 10.1 \\
Inclination Warp   & Yes & $-0.24\pm0.03$ & $479\pm3$ & $3.78\pm0.01$ & 1.08 & $0.04\pm0.01$\tablenotemark{f} & 9.6 \\
Inclination Warp\tablenotemark{g}    & Yes & --0.191& 473.5 & 3.794 & 1.19 & 0.034\tablenotemark{f} & 9.64 \\
\hline
\end{tabular}
\end{center}
\tablenotetext{a}{Offset with respect to reference maser at 510 km~s$^{-1}$ LSR.}
\tablenotetext{b}{Maximum deviation from Keplerian rotation between innermost and outermost masers, positive representing a flattening in the projected rotation curve.}
\tablenotetext{c}{$\xi$ ($\vrot\propto r^{-\xi}$)}
\tablenotetext{d}{$\rho_c$ ($M_\odot$~pc$^{-3}$)}
\tablenotetext{e}{$\dot{M}/\alpha$ (M$_\odot$~yr$^{-1}$)}
\tablenotetext{f}{$di/dr$ (mas$^{-1}$)}
\tablenotetext{g}{$di/dr$ bootstrapped global model}
\end{table}

\clearpage

\clearpage
\newpage

\clearpage

\end{document}